%% file: main.tex
\documentclass{jpp}
\usepackage{graphicx}
\usepackage{epstopdf, epsfig}

\usepackage{amssymb}
\usepackage{amsmath}
\DeclareMathOperator{\sech}{sech}

\usepackage{color}
\usepackage{subcaption}

\usepackage{verbatim} 

\shorttitle{Temperature gradient driven heat flux closure}
\shortauthor{F. Allmann-Rahn, T. Trost and R. Grauer}

\title{Temperature gradient driven heat flux closure in fluid simulations of collisionless reconnection}

\author{F. Allmann-Rahn\aff{1}, T. Trost\aff{1} \and R. Grauer\aff{1} \corresp{\email{grauer@tp1.rub.de}}}

\affiliation{ \aff{1}Institute for Theoretical Physics I, Ruhr University Bochum, Germany }

\begin{document}

\maketitle

\begin{abstract}
\input{abstract}
\end{abstract}

\input{introduction}
\input{vlasov_fluid}
\input{k0_closure}
\input{setup}
\input{comparison}
\input{better_k0}
\input{laplace}
\input{performance}
\input{conclusion}

\input{acknowledgments}

\bibliographystyle{jpp}

\bibliography{bibliography}

\end{document}

%% file: abstract.tex
Recent efforts to include kinetic effects in fluid simulations of plasmas have
been very promising.  Concerning collisionless magnetic reconnection, it has
been found before that damping of the pressure tensor to isotropy leads to good
agreement with kinetic runs in certain scenarios. An accurate representation of
kinetic effects in reconnection was achieved in a study by Wang et al.\ ({\em
Phys. Plasmas}, volume 22, 2015, 012108) with a closure derived from earlier
work by Hammett and Perkins ({\em PRL}, volume 64, 1990, 3019). Here, their approach
is analyzed on the basis of heat flux data from a Vlasov simulation. As a
result, we propose a new local closure in which heat flux is driven by
temperature gradients. That way, a more realistic approximation of Landau
damping in the collisionless regime is achieved. Previous issues are addressed
and the agreement with kinetic simulations in different reconnection setups is
improved significantly. To the authors' knowledge, the new fluid model is the
first to perform well in simulations of the coalescence of large magnetic
islands.

%% file: introduction.tex
\section{Introduction \label{sec:introduction}}

Magnetic reconnection is a process where magnetic field line topology changes
(field lines reconnect) to an energetically more advantageous state. Magnetic
energy is converted into heating and particle acceleration. Reconnection occurs
throughout the universe, e.g.\ in the context of gamma ray bursts, in stellar
and especially solar flares or in Earth's magnetosphere.

Plasma phenomena that happen on large time and spatial scales and those where
collisions are an important factor can often be described sufficiently with
hydrodynamic or fluid models. In many cases, such as collisionless magnetic
reconnection and collisionless shocks, these conditions are not fulfilled and
thus kinetic effects have to be taken into account. However, kinetic simulations
are computationally expensive and problems with large system sizes like
reconnection in the magnetotail or three-dimensional reconnection cannot be
computed with a fully kinetic model. The fluid equations on the other hand can
be orders of magnitude cheaper to compute and can be a good approximation
depending on how well the corresponding closure suits the problem.

\citet{wang-hakim-etal:2015} suggested a heat flux closure which approximates a spectrum of wave numbers by one single
wave number $k_0$. The closure, although simple, gave very good results in fluid simulations of collisionless
reconnection. Nevertheless, Wang et al.\ asserted that further work is needed to improve the closure, e.g.\ by finding
a more suitable $k_0$. One way to do this is to compare the closure approximation to the actual heat flux gained
from a kinetic simulation. This is difficult with a particle in cell (PIC) code because higher moments like pressure
and especially heat flux are very noisy in PIC simulations. We analyze the closure making use of kinetic data from a Vlasov
simulation. Dependence of $k_0$ on plasma parameters is sought as well as other major potential improvements to the closure.

%% file: vlasov_fluid.tex
\section{Vlasov equation and ten moment fluid equations \label{sec:vlasov_fluid}}

A plasma may be described by distribution functions $f_{s}(\mathbf{x}, \mathbf{v}, t)$ which determine the particle density at point
$(\mathbf{x}, \mathbf{v})$ in phase space at time $t$ for the particle species $s$.
Under the assumption that there are no collisions (which is a good approximation e.g.\ for plasmas in space physics),
the evolution of the distribution function is given by the continuity equation
\begin{equation} \frac{\partial f_{s}}{\partial t} + \nabla \cdot (\mathbf{v} f_{s}) + \nabla_{v} \cdot (\mathbf{a} f_{s}) = 0. \end{equation}

Inserting Lorentz acceleration $\mathbf{a} = \frac{q}{m} (\mathbf{E + v \times B})$, the equation can be rearranged to
give the \textbf{Vlasov equation}
\begin{equation}
\frac{\partial f_{s}}{\partial t} +\mathbf{v} \cdot \nabla f_{s} + \frac{q_{s}}{m_{s}} (\mathbf{E + v \times B}) \cdot \nabla_{v} f_{s} = 0.
\label{eq:vlasov} \end{equation}\\
Evolution of electric and magnetic fields is given by Maxwell's equations $\nabla \cdot \mathbf{E} = \frac{\rho}{\epsilon_{0}}$,
$\nabla \cdot \mathbf{B} = 0$, $\nabla \times \mathbf{E} = - \frac{\partial \mathbf{B}}{\partial t}$ and
$\nabla \times \mathbf{B} = \mu_{0} \mathbf{j} + \mu_{0} \epsilon_{0} \frac{\partial \mathbf{E}}{\partial t}$.\\

The charge and current densities are defined as $\rho = \sum_{s} q_{s} n_{s}$ and
$\mathbf{j} = \sum_{s} q_{s} \mathbf{u}_{s}$.
Fluid quantities can be derived by taking moments of the distribution function, i.e.\ multiplying $f_s$ by
powers of $v$ and taking the integral over velocity space. The zeroth moment is the particle density
$ n_s(\mathbf{x}, t) = \int f_s(\mathbf{x}, \mathbf{v}, t) \text{d}\mathbf{v}$. Similarly, the first moment is the mean velocity
$ \mathbf{u}_s(\mathbf{x},t) = \frac{1}{n_s(\mathbf{x}, t)} \int \mathbf{v} f_s(\mathbf{x}, \mathbf{v}, t) \text{d}\mathbf{v}$.
Higher moments include pressure $\text{P}_{s} = m_{s} \int \mathbf{v}' \otimes \mathbf{v}' f_{s} \text{d}\mathbf{v}$
and heat flux $\text{Q}_{s} = \frac{m_{s}}{2} \int \mathbf{v}' \otimes \mathbf{v}' \otimes \mathbf{v}' f_{s} \text{d}\mathbf{v}$,
where $\mathbf{v}' = \mathbf{v - u}$.\\

By taking moments of the whole Vlasov equation, one can obtain the fluid equations. Due to the $\mathbf{v} \cdot \nabla f$ term in the
Vlasov equation, however, every moment contains a quantity that is defined by the next higher moment. Therefore, the resulting system of equations
needs a closure in order to be self consistent. Usually this is done by finding an approximation for pressure or heat flux.
Two common versions of the fluid equations are the five moment equations (pressure closure) and ten moment equations (heat flux closure).\\

The following three equations along with Maxwell's equations and a heat flux closure are the complete set of \textbf{ten moment equations}
($D$ denotes the dimensionality):

\begin{equation}
\frac{\partial n_{s}}{\partial t} + \nabla \cdot (n_{s} \mathbf{u}_{s}) = 0 \, ,
\label{eq:tenmoment_continuity} \end{equation}

\begin{equation}
m_{s} n_{s} (\frac{\partial \mathbf{u_{s}}}{\partial t} + \mathbf{u}_{s} \cdot \nabla \mathbf{u}_{s}) =
n_{s} q_{s} (\mathbf{E} + \mathbf{u_{s}} \times \mathbf{B}) - \nabla \cdot \text{P}_{s} \, ,
\label{eq:tenmoment_movement} \end{equation}

\begin{equation}
\frac{D}{2} (\frac{\partial \text{P}_{s}}{\partial t} + \mathbf{u}_{s} \cdot \nabla \text{P}_{s}) +
\frac{2+D}{2} \text{P}_{s} \nabla \cdot \mathbf{u_{s}} = - \nabla \cdot \text{Q}_{s} \, .
\label{eq:tenmoment_energy} \end{equation}

%% file: k0_closure.tex
\section{WHBG physical space fluid closure for Landau damping \label{sec:k0_closure}}

Many heat flux closures for collisionless plasmas exist and have been successfully applied,
e.g.\ the Landau damping closures by \citet{hammett-perkins:1990} or \citet{passot-sulem:2003}. An overview is given in \citet{chust-belmont:2006}.
Most closures are not designed for heat flux and pressure tensors in three dimensions though.
\citet{hammett-perkins:1990} (also: \citet{hammett-dorland-perkins:1992}) approximated
the plasma response function using a Pade series in order to include Landau damping in the fluid equations
which is the main damping mechanism -- and thus cause of heat flux -- in collisionless plasmas.
The closure was found to be an excellent approximation in many different cases (see e.g.\ the study by \citet{sarazin-dif-pradalier-etal:2009}).\\

The Hammett-Perkins closure is given in one-dimensional Fourier space as
\begin{equation}
\tilde{q}_{k} = - n_{0} \chi_{1} \frac{2^{1/2} v_{t}}{|k|} i k \tilde{T}_{k} \label{eq:hammett_perkins}
\end{equation}
with  $v_t = \sqrt{k_{B} T / m}$ and $\chi_{1} = 2/\sqrt{\pi}$.
The closure resembles Fick's second law $q = - n D \frac{\partial T}{\partial x}$.\\

It was found by \cite{johnson-rossmanith:2010} that heat flux in collisionless reconnection can be modeled
by a relaxation of the pressure tensor to an isotropic equilibrium pressure. This can be motivated with the
Hammett-Perkins closure: Eq. \ref{eq:hammett_perkins} was simplified by \citet{wang-hakim-etal:2015} in order
to be applicable in three dimensional physical space. Since $\nabla \cdot \mathrm{Q}$ shall be approximated, the divergence of
Eq.\ \ref{eq:hammett_perkins} is taken which gives
\begin{equation}
    \mathrm{LHS} = i\ \mathbf{k}^{\mathrm{t}} \cdot \mathrm{Q}_{s} 
    \label{eq:generalization_part1}
\end{equation}
on one side and
\begin{equation}
    \mathrm{RHS} = n_0 \chi_{1} \frac{2^{1/2} v_{t}}{|k|}\ \mathbf{k}^{\mathrm{t}}  \cdot (\mathbf{k}^{\mathrm{t}} \cdot \mathrm{\tilde{T}}_{s})
    \label{eq:generalization_part2}
\end{equation}
on the other side of the equation. Here, $\mathbf{k}^{\mathrm{t}}$ is the transposed wave vector.
It becomes obvious that a direct generalization of Fick's law to tensors is not possible
since Eq.\ \ref{eq:generalization_part1} is a second-order tensor and Eq.\ \ref{eq:generalization_part2} is a scalar. Therefore, the vector character
of $\mathbf{k}$ was neglected on the right-hand side (and the constant $\chi_{1} \sqrt{2} \approx 1.6$ was dropped), resulting in
\begin{equation}
  i k_{m} \mathrm{Q}_{ijm}(k) \approx n_{0} \frac{v_{t}}{|k|} k^{2}\ \tilde{\mathrm{T}}_{ij}(k)
  = n_{0} v_{t} |k|\ \tilde{\mathrm{T}}_{ij}(k) \, .
  \label{eq:k0_derivation}
\end{equation}

The adjustment done by treating $\mathbf{k}$ as a scalar is that (in physical space) $\nabla \cdot (\nabla \cdot \mathrm{T}_{s})$ is
replaced by $\nabla^{2}\ \mathrm{T}_{s}$, i.e.\ the Laplace operator is used on each component of $\mathrm{T}_{s}$. At the same time
regular divergence is taken on the left-hand side. A motivation for this approximation is given in Sec.\ \ref{sec:laplace}.\\

The perturbed temperature $\tilde{\mathrm{T}}_{ij}$ can be expressed as $(\mathrm{P}_{ij} - p \delta_{ij})/n_{0}$, where
$p \delta_{ij}$ is the isotropic pressure with $p = (\mathrm{P}_{xx} + \mathrm{P}_{yy} + \mathrm{P}_{zz})/3$. Thus
\begin{equation} i k_{m} \mathrm{Q}_{ijm}(k) \approx v_{t}\ |k|\ (\mathrm{P}_{ij}(k) - p(k) \delta_{ij}). \end{equation}

Finally, the wave number field $k$ is approximated by one single wave number $k_{0}$, so that the closure
can be written in physical space as
\begin{equation}
\partial_{m} \mathrm{Q}_{ijm} \approx v_{t}\ |k_{0}|\ (\mathrm{P}_{ij} - p \delta_{ij}).
\label{eq:wang-hakim-etal:2015} \end{equation}
We will refer to this closure as the scalar-k closure in this paper.\\

%% file: setup.tex
\section{Numerical setup\label{sec:setup}}

Fluid and kinetic Vlasov simulations of different reconnection problems are performed. The Vlasov
code is described in \citet{schmitz-grauer:2006b,schmitz-grauer:2006}, the fluid code and its coupling to the Vlasov code is presented in \citet{rieke-trost-grauer:2015}. Time is normalized over
inverse ion cyclotron frequency $\mathrm{\Omega_{i,0}^{-1}}$, length over ion inertial length $d_{i,0}$,
speed over Alfv\'en velocity $v_{A,0}$ and mass over ion mass $m_{i}$. The electron-ion mass ratio is $m_{i}/m_{e} = 25$
in all simulations.

\subsection{GEM}

The GEM (Geospace Environmental Modeling) reconnection setup \citep{birn-drake-shay-etal:2001} is a reconnection problem that uses
a Harris sheet configuration \citep{harris:1962}. The initial magnetic field is given by
$B_{x}(y) = B_{0} \tanh(y/\lambda)$ and the particle density by $n(y) = n_{0} \sech^{2}(y/\lambda) + n_{b}$
where $\lambda = 0.5$, $B_{0} = 1, n_{0} = 1$ and the background density $n_{b} = 0.2$. Temperature is defined by $n_{0}(T_{e}+T_{i}) = B_{0}^{2}$,
$T_{i}/T_{e} = 5$. Speed of light is set to $c = 20\ v_{A,0}$. The domain is of size $L_{x} \times L_{y} = (8\pi\times4\pi)\ d_{i,0}$.
It is translationally symmetric in $z$-direction, periodic in $x$-direction
and has conducting walls for fields and reflecting walls for particles in $y$-direction. In order to start the reconnection process, a perturbation
in the magnetic field is applied that takes the form $\mathbf{B} = \hat{\mathbf{z}} \times \nabla \psi$ where the perturbation in the magnetic flux
is given by $\psi(x,y) = 0.1\, \cos(2 \pi x / L_{x})\, \cos(\pi y / L_{y})$. Because of symmetries, it is sufficient to simulate one fourth of the domain.
The time span covered by the Vlasov simulation is $40\ \mathrm{\Omega_{i,0}^{-1}}$, reconnection rate peaks at $t \approx 20\ \mathrm{\Omega_{i,0}^{-1}}$.
The domain is resolved by $256 \times 128$ cells.

\subsection{Large Harris sheet -- WHBG}

Reconnection in Earth's magnetotail happens on much larger spatial scales than reconnection in the GEM setup. In order to approach
larger scales, \citet{wang-hakim-etal:2015} performed kinetic and fluid simulations in a configuration like
GEM but with a $(100 \times 50)\ d_{i,0}$ domain and $c = 15\ v_{A,0}$. For simple reference, it will be called the WHBG setup
in this paper. A study of reconnection in a domain of this size was done before by \citet{daughton-scudder-karimabadi:2006}, but with
open boundary conditions unlike the WHBG version.

\subsection{Island coalescence}

The island coalescence reconnection problem has also been studied extensively, e.g.\ by \citet{karimabadi-dorelli-etal:2011} (large PIC simulations),
\citet{stanier-daughton-etal:2015} (PIC, hybrid and Hall-MHD compared) and \citet{ng-huang-hakim-etal:2015} (MHD, Hall-MHD and ten moment fluid simulations).
We use the same parameters as the aforementioned studies. The initial configuration is a Fadeev equilibrium \citep{fadeev-kvabtskhava-komarov:1965}:
$A_{z} = - \lambda B_{0} \ln( \cosh(y/\lambda) + \epsilon \cos(x/\lambda)$ and
$n = n_{0} (1 - \epsilon^{2}) / (\cosh(y/\lambda) + \epsilon \cos(x/\lambda))^{2} + n_{b}$ with $\epsilon = 0.4$, $n_{b} = 0.2$ and a variable $\lambda$.
Temperature is $T = T_{i} = T_{e} = 0.5$, speed of light $c = 15\ v_{A,0}$ and the domain size is proportional to $\lambda$ according to
$L_{x} \times L_{y} = (2 \pi \lambda \times 4 \pi \lambda)\ d_{i,0}$. The boundaries are periodic in $y$-direction and conducting for fields and
reflecting for particles in $x$-direction. The B-field perturbation is
$ \delta B_{x} = 0.1\, \sin(y/(2 \lambda) - \pi)\, \cos(x/(2 \lambda))$ and $\delta B_{y} = -0.1\, \cos(y/(2 \lambda) - \pi)\, \sin(x/(2 \lambda))$
\citep{daughton-roytershteyn-etal:2009}. Time is normalized to Alfv\'en time $t_{A} = L_{y}/v_{A,0}$. The normalized reconnection rate $E_{R}$ is computed
as $E_{R} = \frac{\partial \Psi}{\partial t} / (B' v'_{A})$ where $B'$ is the maximum of the absolute value of the magnetic field between the X-point
and the O-point at $x = 0$ and $t = 0$ and $v'_{A} = B'/\sqrt{\mu_{0} n_{0} m_{i}}$. The magnetic flux is $\Psi = \int^{\mathrm{X}}_{\mathrm{O}}\ \mathrm{dy}\ B_{x}$
(integral from the O- to the X-point).

%% file: comparison.tex
\section{Comparison of heat flux data and the scalar-k closure\label{sec:comparison}}

In order to examine how well the actual divergence of heat flux
agrees with the closure approximation, both sides of Eq.\ \ref{eq:wang-hakim-etal:2015} have been computed. While
Wang et al.\ chose $1/d_{e,0}$ as $k_{0}$ for all components, ideally one can find a better $k_{0}$
by analyzing the plots (cf.\ Sec.\ \ref{sec:better_k0}) of the kinetic simulations. The comparison is done with simulations
of the GEM setup.\\

\begin{figure}
\begin{minipage}{0.365\textwidth} \includegraphics[height=2.15cm]{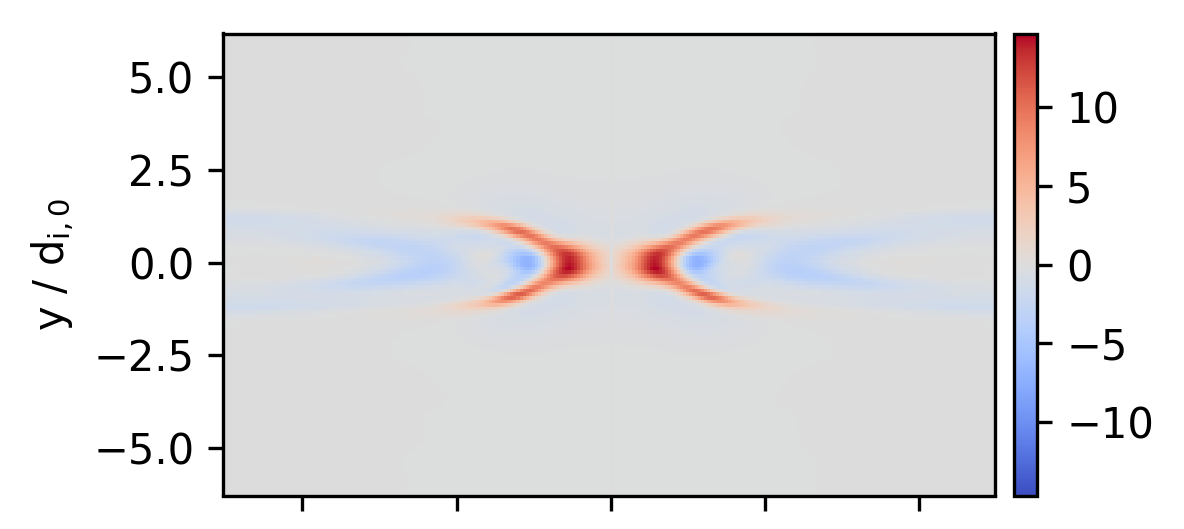} \end{minipage}
\begin{minipage}{0.305\textwidth} \includegraphics[height=2.15cm]{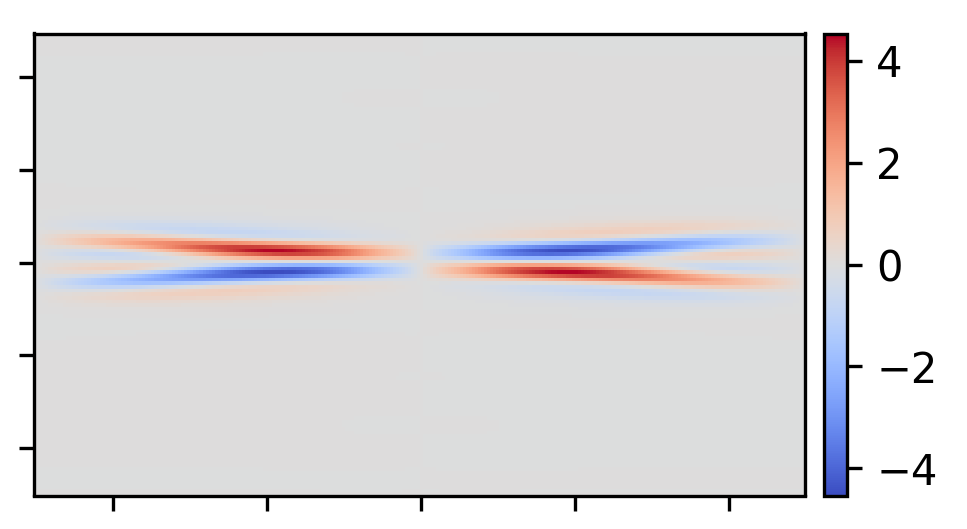} \end{minipage}
\begin{minipage}{0.305\textwidth} \includegraphics[height=2.15cm]{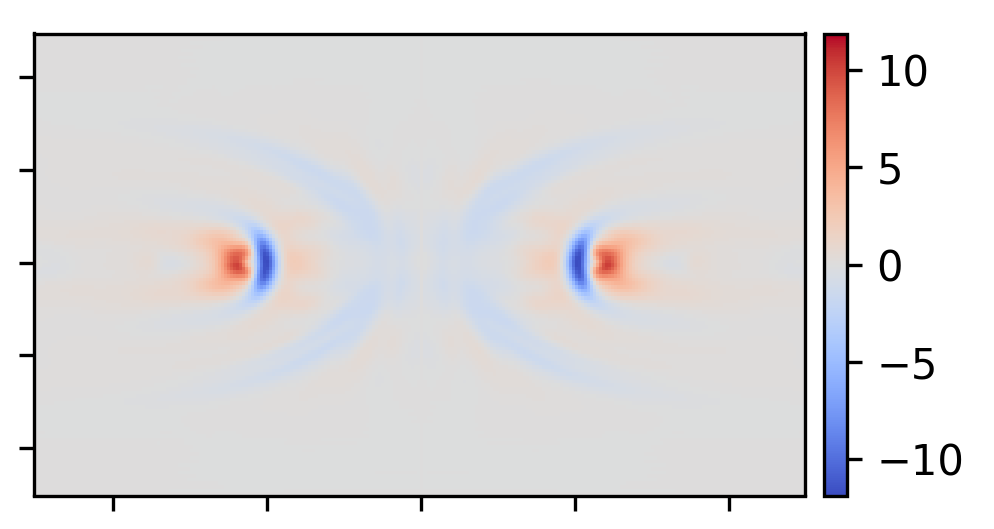} \end{minipage}\\
\begin{minipage}{0.365\textwidth} \includegraphics[height=2.69cm]{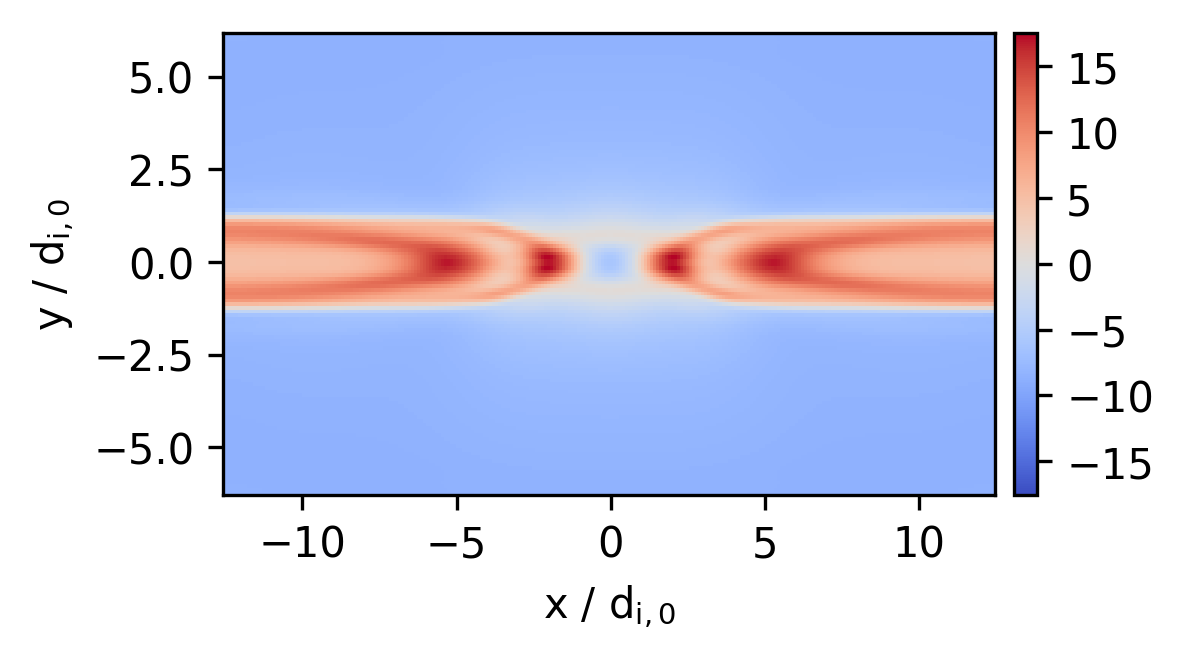} \subcaption{}\end{minipage}
\begin{minipage}{0.305\textwidth} \includegraphics[height=2.69cm]{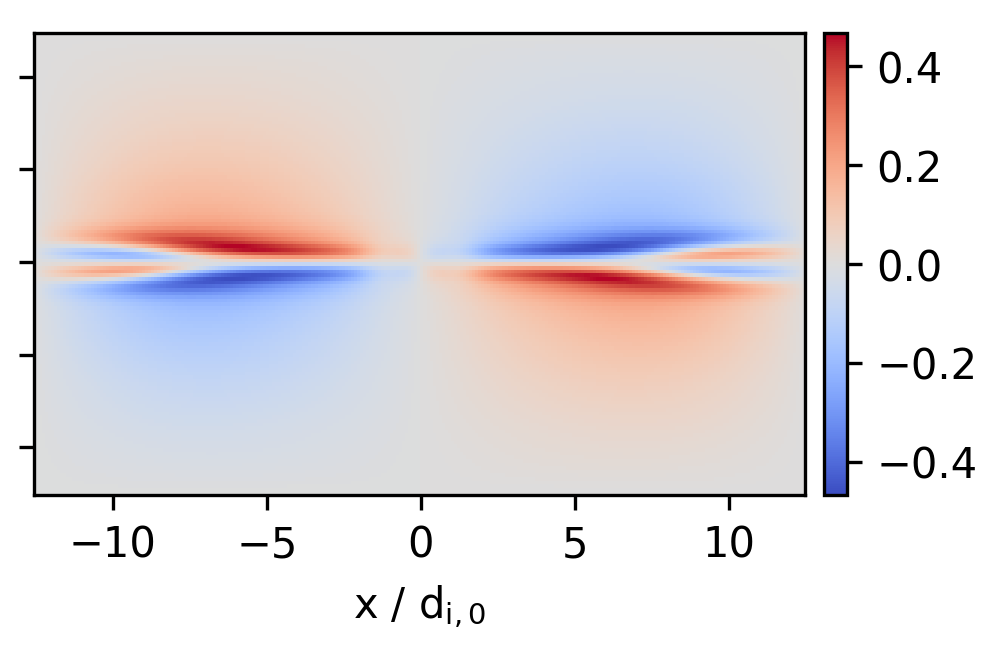} \subcaption{}\end{minipage}
\begin{minipage}{0.305\textwidth} \includegraphics[height=2.69cm]{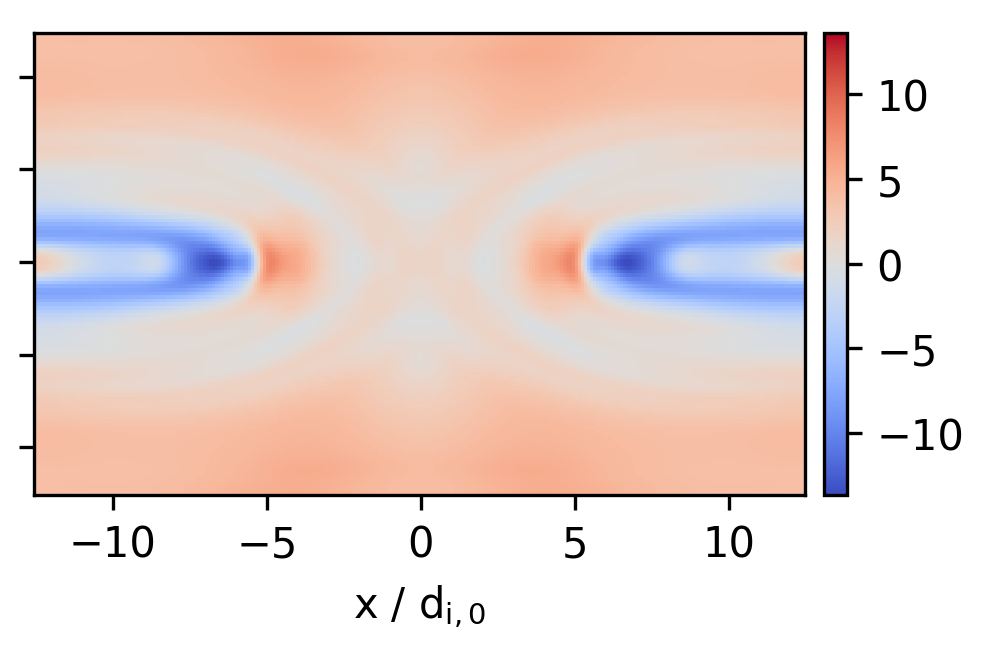} \subcaption{}\end{minipage}

\caption{Comparison of actual heat flux change (first row) and scalar-$k$ closure (second row).
(a) $(\nabla \cdot \mathrm{Q_{e}})_{\mathrm{xx}}$ at t = $17.5\ \mathrm{\Omega_{i}^{-1}}$,
(b) $(\nabla \cdot \mathrm{Q_{e}})_{\mathrm{xy}}$ at t = $7.5\ \mathrm{\Omega_{i}^{-1}}$,
(c) $(\nabla \cdot \mathrm{Q_{e}})_{\mathrm{zz}}$ at t = $30\ \mathrm{\Omega_{i}^{-1}}$.}

\label{fig:comp_k0}\end{figure}

Taking symmetry into account, the heat flux tensor $\mathrm{Q}_{s}$ has ten and the pressure tensor $\mathrm{P}_{s}$ has six independent components.
Therefore Eq.\ \ref{eq:wang-hakim-etal:2015} results in six separate equations, one for each of the pressure tensor's
components.
$1/d_{e,0}$ equates to $5\ d_{i,0}^{-1}$
since the electron-ion mass ratio $m_{e}/m_{i}$ is $1/25$. For the purpose of comparison, a value
of $k_{0} = 1\ d_{i,0}^{-1}$ is used to compute the closure. Representative plots are shown in Fig.\ \ref{fig:comp_k0}.\\

Overall the agreement is decent considering that heat flux often has a complex shape. The approximation is
best in the period before and during reconnection. In the beginning of the simulation the magnitude is usually
off by a factor between $\frac{1}{10}$ and $10$ whereas after reconnection the shape of heat flux generally
becomes very convoluted which is hard to replicate with a closure.

Fig.\ \ref{fig:comp_k0}a shows typical issues insofar as the basic structure is approximated well (the area in
the center of the plot) but other parts of the shape are wrong (here the outer area around the x-axis). Another
recurring problem is that the whole outer region usually has no heat flux, which is wrongly predicted by the
scalar-k closure. A major improvement with correct damping in this outer region will be presented in Sec.\ \ref{sec:laplace}.

A positive example of the scalar-k closure is given by Fig.\ \ref{fig:comp_k0}b, showing how it can even cover details
like the changing sign at the left and right border around the x-axis.
After reconnection, structures tend to get complicated (Fig.\ \ref{fig:comp_k0}c) and while the shape is still
similar, both sign and the location of extrema are inaccurate. This holds true for many components towards
the end of the simulation (35 to 40 $\mathrm{\Omega_{i,0}^{-1}}$).
Concerning a fixed $k_{0}$, the comparison suggests that values between 0.1 and 10 $d_{i,0}^{-1}$ can be reasonable choices.

%% file: better_k0.tex
\section{Searching for a parameter dependent $k_0$\label{sec:better_k0}}

The field of wave numbers $k$ from Eq.\ \ref{eq:hammett_perkins} was replaced with a single fixed number $k_{0}$
by \citet{wang-hakim-etal:2015}. Although this is a massive simplification, it already leads to good results. Nonetheless, there are
differences between a kinetic simulation and a ten moment fluid simulation using the scalar-$k$ closure. Discrepancies exist e.g.\ in the
pressure tensor which may be attributed to the issue that there is no trivial way to generalize the original Hammett-Perkins
closure to three dimensions and that $k$ is the same for each component of the pressure tensor.\\

The idea behind the approximation is that a $k_{0}$ represents the average length scales at which Landau damping occurs in the
given scenario. \cite{wang-hakim-etal:2015} found $k_{0} = 1/d_{e,0} = 5\ d_{i,0}^{-1}$ to fit well in their $(100 \times 50)\ d_{i,0}$
reconnection setup. Tests showed that in the original GEM reconnection problem, however, $k_{0,i} = 0.3\ d_{i,0}^{-1}$ seems to be the optimal value.
This is unintuitive as usually a smaller domain size would not require a smaller (especially much smaller) characteristic wave number.
That means $k_{0}$ seems to be specific to the respective problem. It would be desirable to find a consistent,
variable $k_{0}$ which might depend on local plasma parameters.

Eligible plasma parameters were investigated experimentally by computing the closure with the respective
$k_{0}$ candidate and comparing it to the actual divergence of heat flux as done in Sec.\ \ref{sec:comparison}. Promising
candidates were additionally tested in a ten moment fluid simulation which was then compared to a run with $k_{0} = 5\ d_{i,0}^{-1}$
and a Vlasov run.
Dependence of $k_{0}$ on plasma parameters and quantities was examined in the GEM setup, but none of the experiments
lead to an improvement. This is because the closure's shape is
overall decent and differences to the actual heat flux are often very complex or vary heavily in time and component.
Results from the analysis done in Sec.\ \ref{sec:comparison} and in this section suggest that the deficiencies concerning
shape and sign cannot be fixed with a scalar $k_{0}$ dependent on plasma parameters. The same applies to the wrong
magnitudes that appear early in the simulation because they don't relate to the fluid quantities.

%% file: laplace.tex
\section{Modified, gradient driven closure\label{sec:laplace}}

The Hammett-Perkins closure was transferred to physical space because a Fourier space representation may be computationally expensive
in a physical space code. More important is the issue that it is not clear how to generalize the closure to three-dimensional
tensors. A generalization to tensors in Fourier space was proposed by \citet{ng-hakim-etal:2017}.
They started with Eq.\ \ref{eq:hammett_perkins} and searched for a total
symmetric generalisation of the heat flux $Q_{ijm}$ resulting in
\begin{equation}
  Q_{ijm}(\mathbf{x}) = n(\mathbf{x}) \tilde{Q}_{ijm}(\mathbf{x}) \,, \;\;\;
  \hat{\tilde{Q}}_{ijm}(\mathbf{k}) = -i \frac{v_t}{|k|} \chi k_{[i} \hat{T}_{jk]} \, ,
  \label{eq:ng_global}
\end{equation}
where $\hat{\tilde{Q}}_{ijm}$ and $\hat{T}_{jk}$ denote the Fourier transforms of $\tilde{Q}_{ijm}$ and $T_{jk}$.\\

We take a different approach and focus not on the heat flux directly but on its
divergence $\partial_m Q_{ijm}$. To
attain the symmetry of this divergence tensor, \cite{wang-hakim-etal:2015} used
$-k^{2}\ \mathrm{P}$ in place of the derivative $- \mathbf{k}^t \cdot
(\mathbf{k}^t \cdot \mathrm{P})$ and then approximated $k$ by $k_{0}$.
The physical space equivalent of this is to replace $\nabla \cdot (\nabla \cdot
\mathrm{P})$ by $\nabla^{2} \mathrm{P}$ where now the second approximation ($k
\approx k_{0}$) is not needed. This way the dependence on $k_{0}$ is reduced and
different relaxation in each component of the pressure tensor is allowed. The
resulting expression takes the form of a Fick's law  and thus the damping
nature of the Hammett-Perkins closure is clearly retained.

Our candidate for a new collisionless heat flux closure is

\begin{equation}
\partial_{m} \mathrm{Q}_{ijm} = -\frac{v_{t}}{|k_{0}|}\ \nabla^{2}\ (\mathrm{P}_{ij} - p \delta_{ij}),
\label{eq:d2p}\end{equation}
where the symmetry of the divergence of the heat flux appears naturally.
We will call it the gradient closure in this paper.\\

It is yet to be motivated why $-k^{2}\ \mathrm{P}$ might be a suitable approximation of the derivative.
In order to do so, assume $\mathbf{B} = (B_{x}, 0, 0)$. The wave vector $\mathbf{k}$ is related to plasma
oscillations, therefore $\mathbf{k} \parallel \mathbf{B}$ and $\mathbf{k} = (k_{x}, 0, 0)$.
The Hammett-Perkins closure in Wang et al.'s three-dimensional version is

\begin{equation}
i k_{m} \mathrm{Q}_{ijm} =  \chi_{1} \frac{2^{1/2} v_{t}}{|k|} k^{2}\ (\mathrm{P}_{ij} - p \delta_{ij})
\end{equation}

or

\begin{equation}
i k_{x} \mathrm{Q}_{ijx} + i k_{y} \mathrm{Q}_{ijy} + i k_{z} \mathrm{Q}_{ijz} =
\chi_{1} \frac{2^{1/2} v_{t}}{|k|} (k_{x}^{2} + k_{y}^{2} + k_{z}^{2})\ (\mathrm{P}_{ij} - p \delta_{ij}).
\end{equation}\\

After dividing by $i k_{x}$ and with $k_{y} = k_{z} = 0$, the equation has the form of the original Hammett-Perkins closure

\begin{equation}
\mathrm{Q}_{ijx}(k) =  - \chi_{1} \frac{2^{1/2} v_{t}}{|k_{x}|} i k_{x}\ (\mathrm{P}_{ij}(k) - p(k) \delta_{ij}).
\label{eq:hammett_perkins_3d} \end{equation}\\

Hence, Eq.\ \ref{eq:wang-hakim-etal:2015} is a plausible generalization to three-dimensional tensors along magnetic field lines. This
indicates that, while not exact in all space, the simplification of treating $k$ as a scalar is reasonable.\\

\begin{figure}
\begin{minipage}{0.32\textwidth} \includegraphics[height=2.45cm]{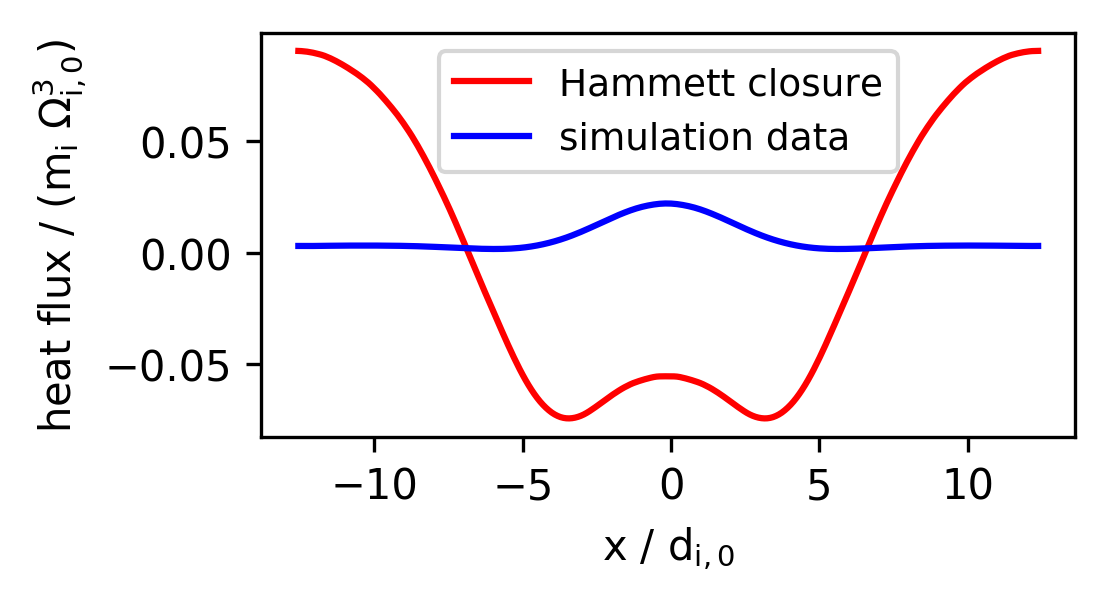} \end{minipage}
\begin{minipage}{0.32\textwidth} \includegraphics[height=2.45cm]{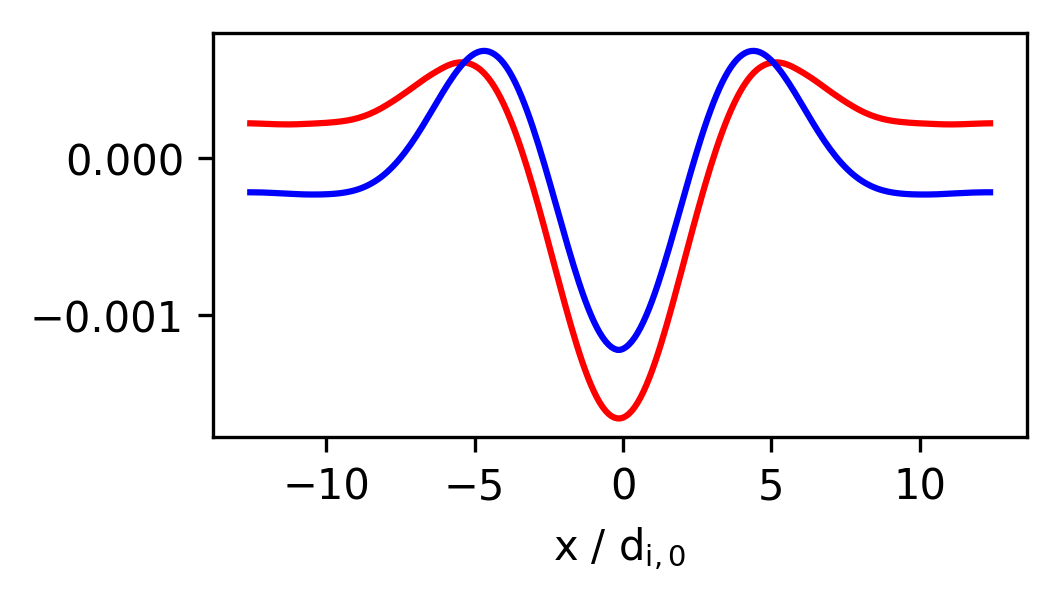} \end{minipage}
\begin{minipage}{0.32\textwidth} \includegraphics[height=2.45cm]{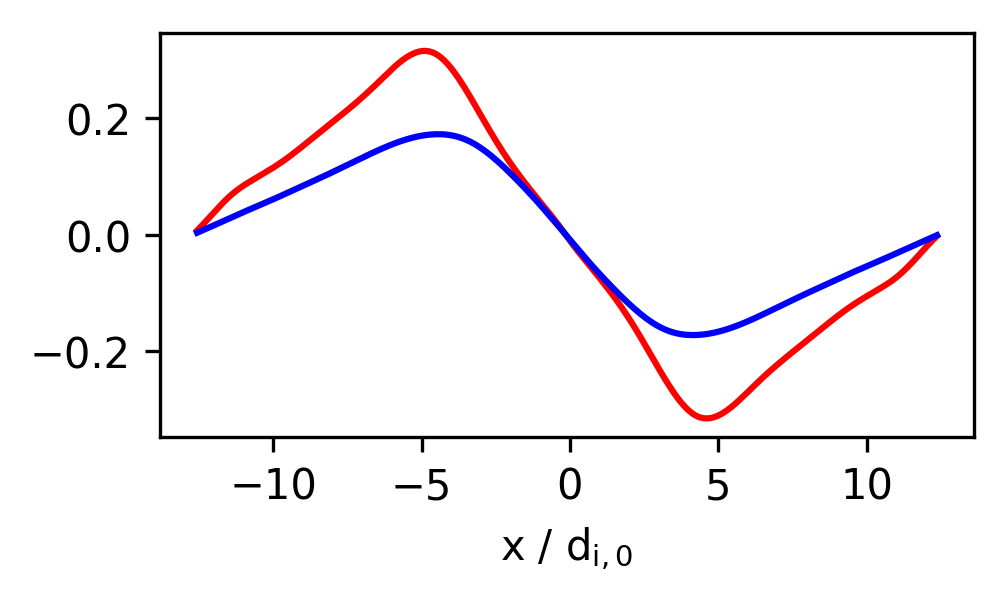} \end{minipage}
\caption{Hammett-Perkins closure, components from left to right: $\mathrm{Q_{xxz,e}}, \mathrm{Q_{xxz,i}}, \mathrm{Q_{xyy,e}}$.\\}
\label{fig:hilbert}\end{figure}

In the outer region of the simulation, field lines are nearly parallel to the x-axis. Thus, it is straightforward to use Eq.\ \ref{eq:hammett_perkins_3d}
to test whether the Hammett-Perkins closure can be applied to the problem of
reconnection or, more precisely, in how far the issues found are related to the choice of $k_{0}$ and the scalar-$k$ approximation
and in how far the problems are inherent to the original closure.
The heat flux according to the closure was calculated by Fourier transforming a one-dimensional section of
the pressure data with $y=const.$, multiplying it by $-i \frac{k}{|k|} = -i\ \mathrm{sgn}(k)$ and then Fourier transforming back to physical space. This
corresponds to the Hilbert transform of the perturbed pressure. Data was taken at $t = 17.5\ \Omega_{i}^{-1}$ and $y = -4.27\ d_{i,0}$.
The results, compared to the actual heat flux, are plotted exemplarily (Fig.\ \ref{fig:hilbert}). Their shape is often correct
but there are deviations from the heat flux data in magnitude or even sign. So the issues are similar to those of the scalar-$k$ closure,
although to lesser extent.\\

\begin{figure}
\begin{minipage}{0.365\textwidth} \includegraphics[height=2.15cm]{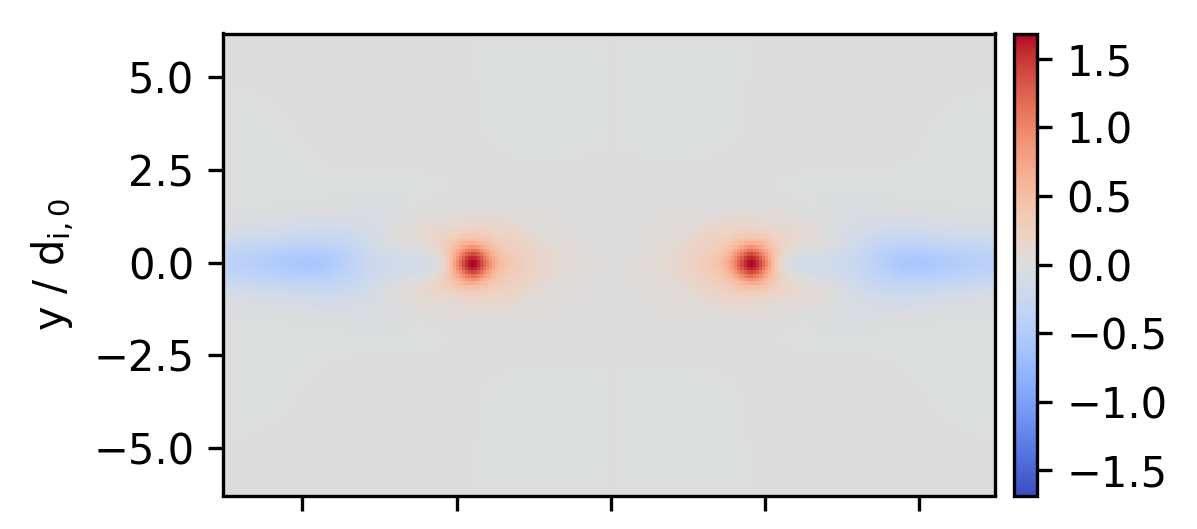} \end{minipage}
\begin{minipage}{0.305\textwidth} \includegraphics[height=2.21cm]{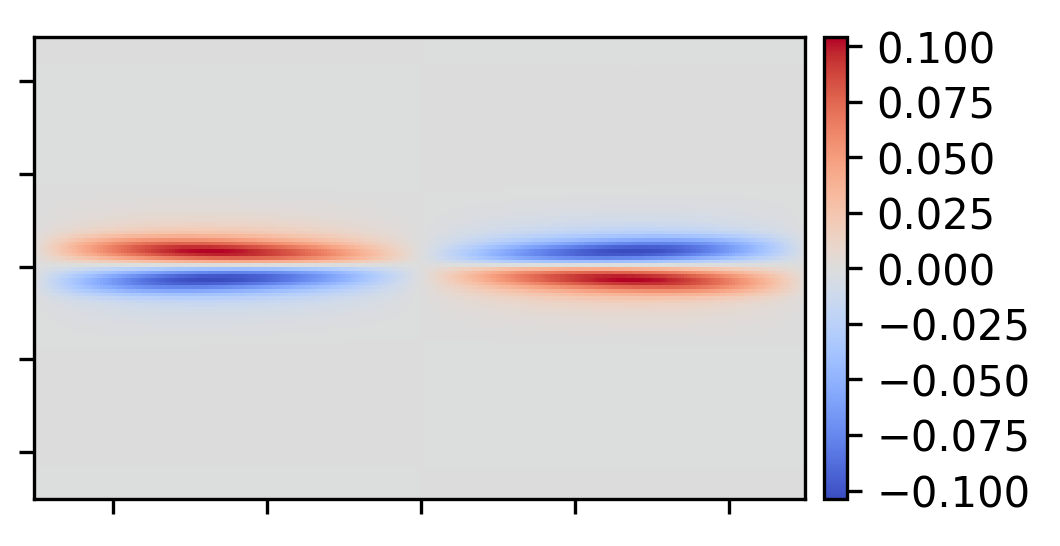} \end{minipage}
\begin{minipage}{0.305\textwidth} \includegraphics[height=2.17cm]{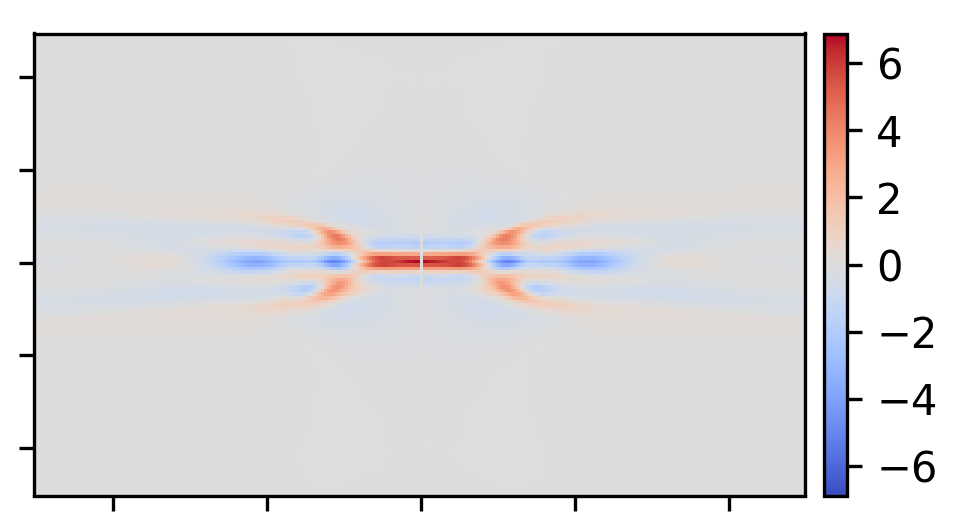} \end{minipage}\\

\begin{minipage}{0.365\textwidth} \includegraphics[height=2.15cm]{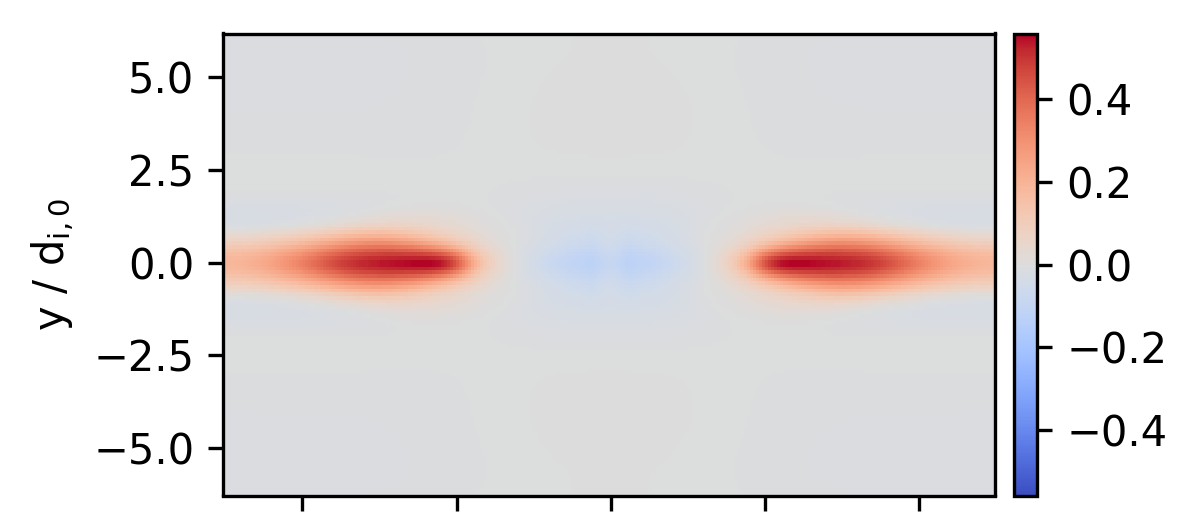} \end{minipage}
\begin{minipage}{0.305\textwidth} \includegraphics[height=2.15cm]{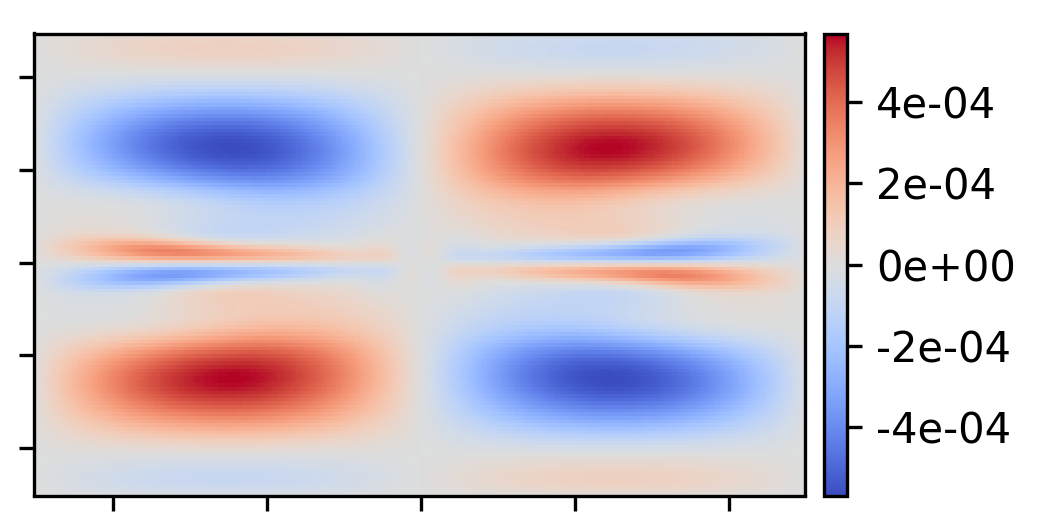} \end{minipage}
\begin{minipage}{0.305\textwidth} \includegraphics[height=2.22cm]{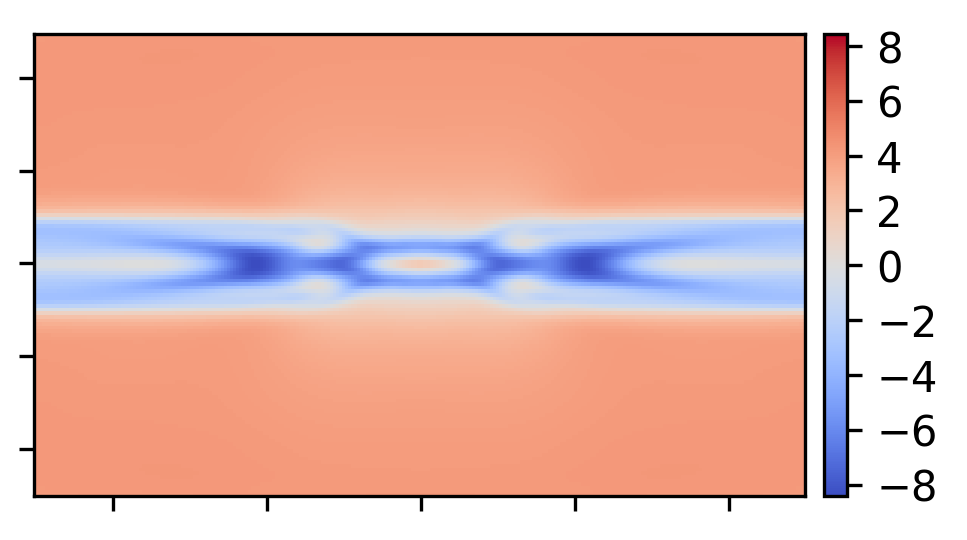} \end{minipage}\\

\begin{minipage}{0.365\textwidth} \includegraphics[height=2.73cm]{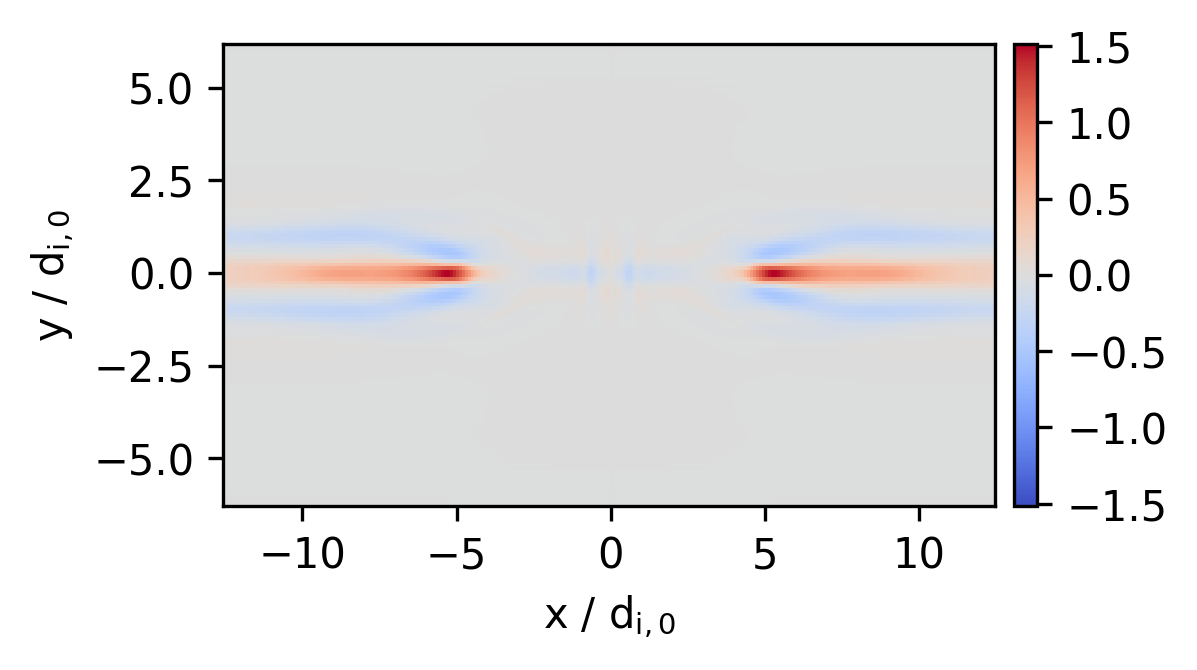} \subcaption{} \end{minipage}
\begin{minipage}{0.305\textwidth} \includegraphics[height=2.73cm]{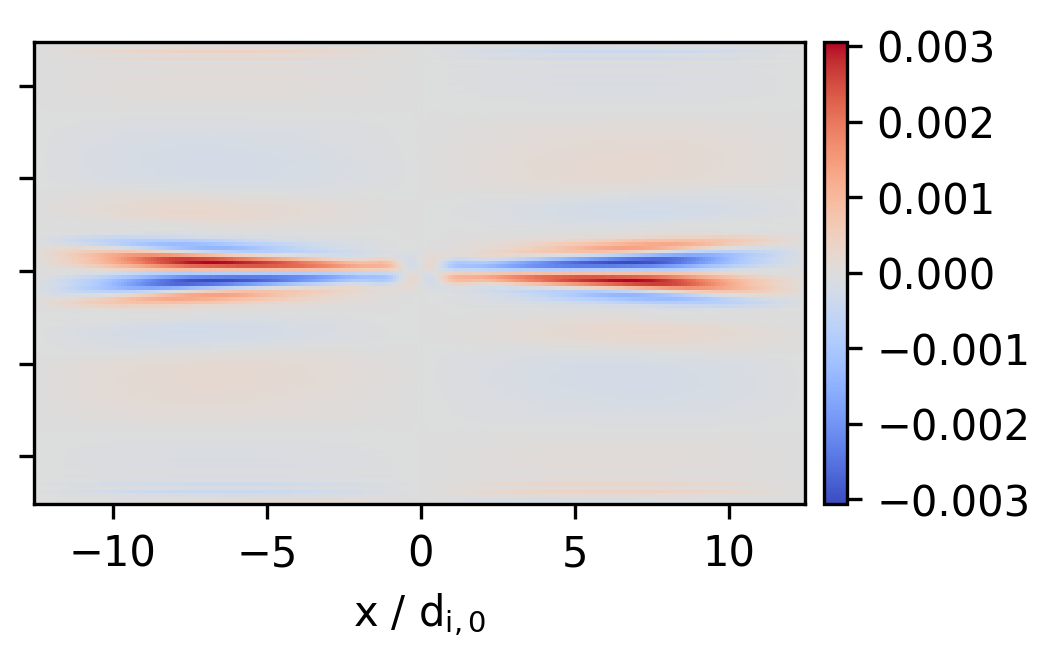} \subcaption{} \end{minipage}
\begin{minipage}{0.305\textwidth} \includegraphics[height=2.73cm]{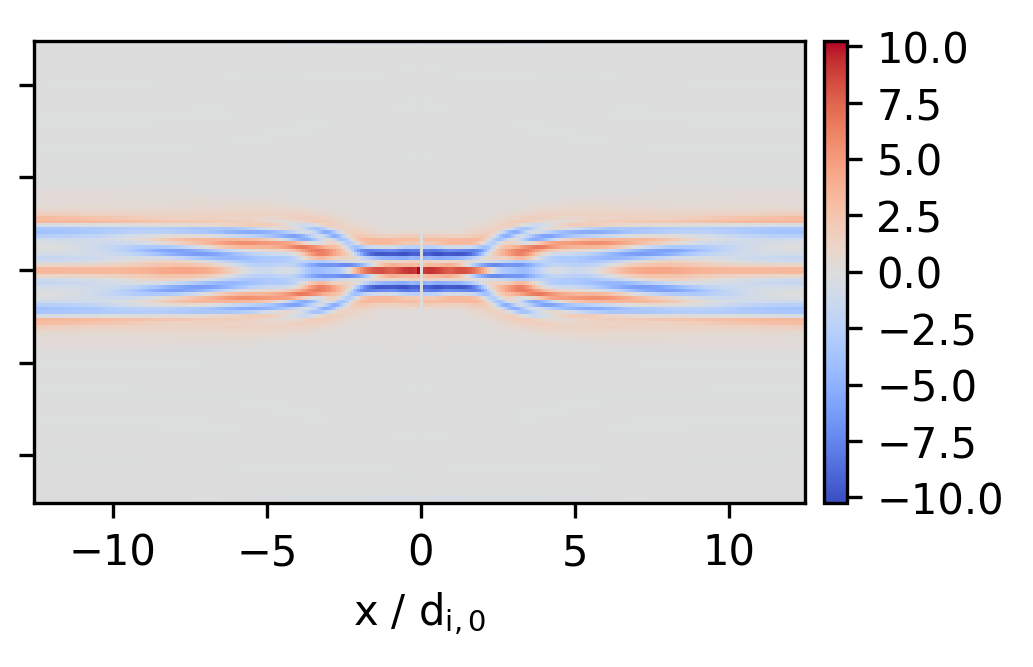} \subcaption{} \end{minipage}

\caption{Comparison of actual heat flux change (first row), scalar-$k$ closure (second row) and gradient closure (third row).
(a) $(\nabla \cdot \mathrm{Q_{i}})_{\mathrm{xx}}$ at t = $20\ \mathrm{\Omega_{i}^{-1}}$,
(b) $(\nabla \cdot \mathrm{Q_{i}})_{\mathrm{xy}}$ at t = $7\ \mathrm{\Omega_{i}^{-1}}$,
(c) $(\nabla \cdot \mathrm{Q_{e}})_{\mathrm{yy}}$ at t = $17.5\ \mathrm{\Omega_{i}^{-1}}$.}
\label{fig:comp_lp}\end{figure}

As done before with the scalar-$k$ closure, we also compare the gradient closure to heat flux data from a Vlasov run.
A comparison of magnitude suggests $k_{0,s} = 3/d_{s,0}$ for the characteristic wave number in the new closure.
This is also the value that was used for the plots in Fig.\ \ref{fig:comp_lp}.
Improvements are recognizable like a better representation of extrema (Fig.\ \ref{fig:comp_lp}a). It has been asserted before that the scalar-$k$ closure
sometimes yields bad results in the outer regions which was not the case with the original Hammett-Perkins closure. This issue has
indeed been fixed using the Laplacian as can be seen in Fig.\ \ref{fig:comp_lp}b and \ref{fig:comp_lp}c. Recent efforts to
couple the Vlasov equation to ten moment fluid equations \citep{rieke-trost-grauer:2015,trost-lautenbach-grauer:2017} could profit from the improvement in the outer
region since that is where the fluid model would be used in a coupling scenario.\\

%% file: performance.tex
\section{The pressure gradient closure in reconnection simulations\label{sec:performance}}

\subsection{GEM}

\begin{figure}
\begin{minipage}{0.365\textwidth} \includegraphics[height=2.15cm]{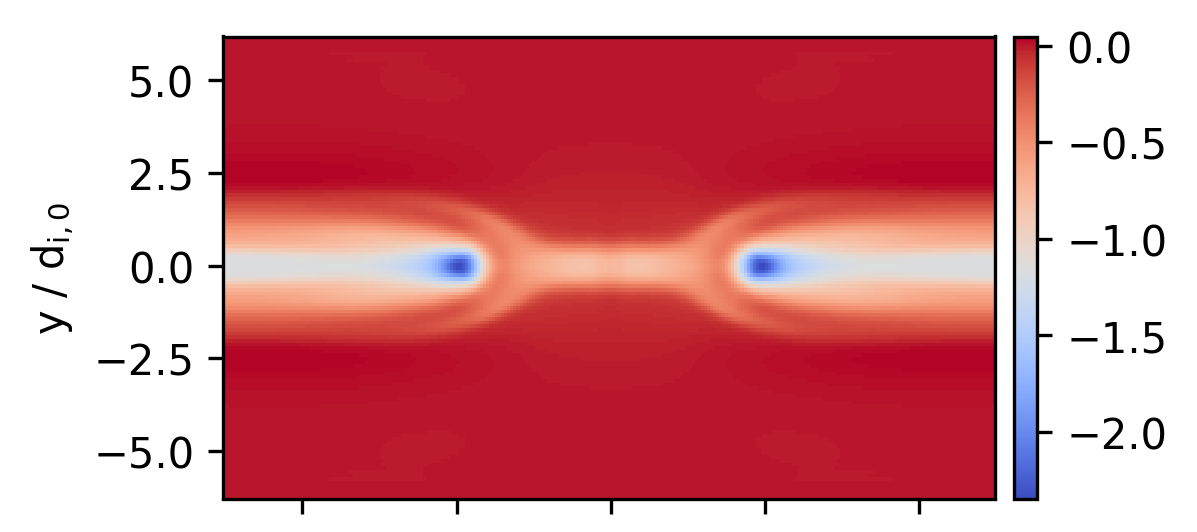} \end{minipage}
\begin{minipage}{0.305\textwidth} \includegraphics[height=2.15cm]{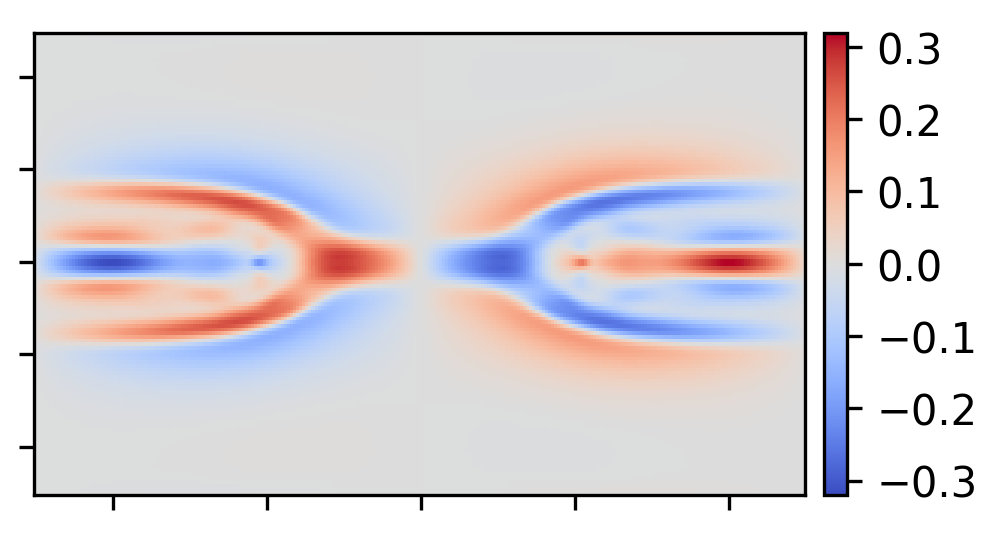} \end{minipage}
\begin{minipage}{0.305\textwidth} \includegraphics[height=2.15cm]{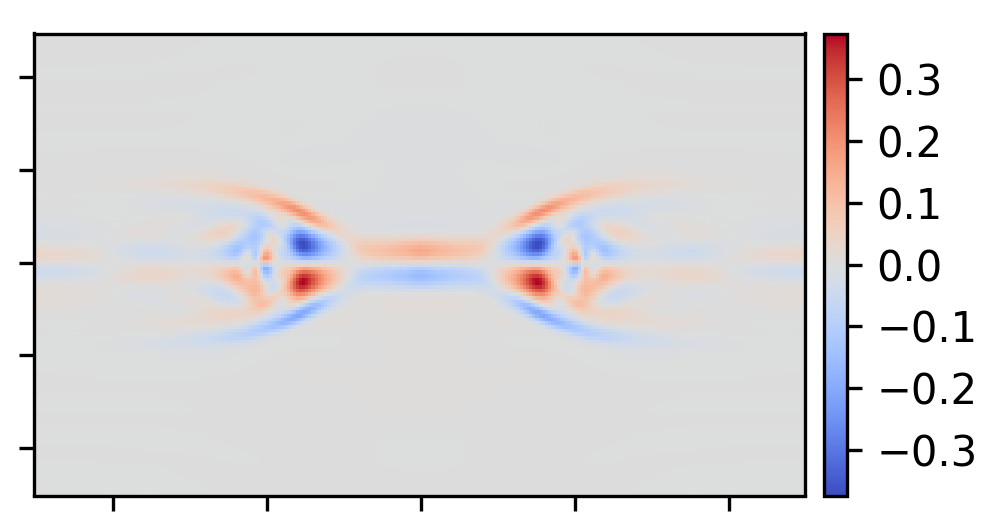} \end{minipage}\\

\begin{minipage}{0.365\textwidth} \includegraphics[height=2.15cm]{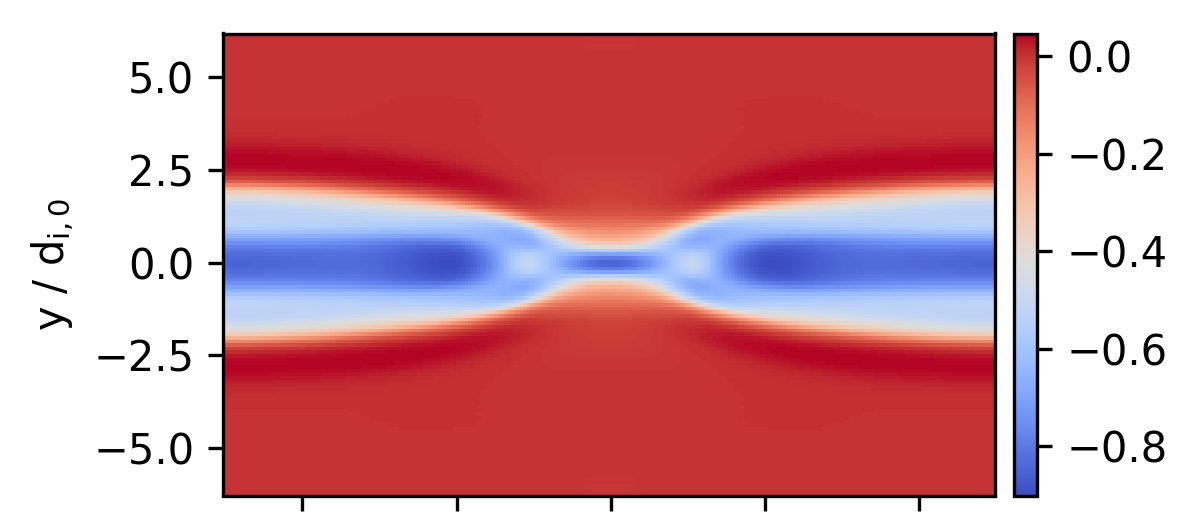} \end{minipage}
\begin{minipage}{0.305\textwidth} \includegraphics[height=2.15cm]{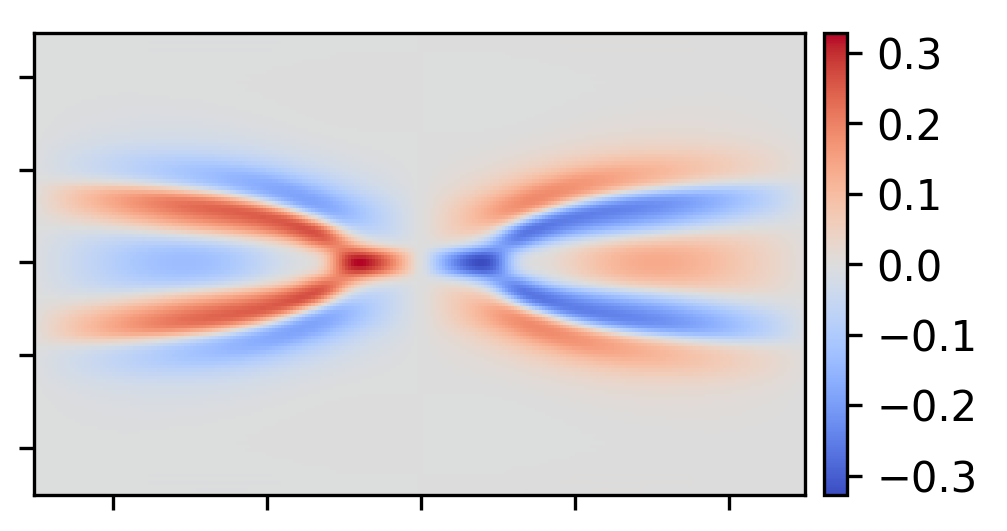} \end{minipage}
\begin{minipage}{0.305\textwidth} \includegraphics[height=2.15cm]{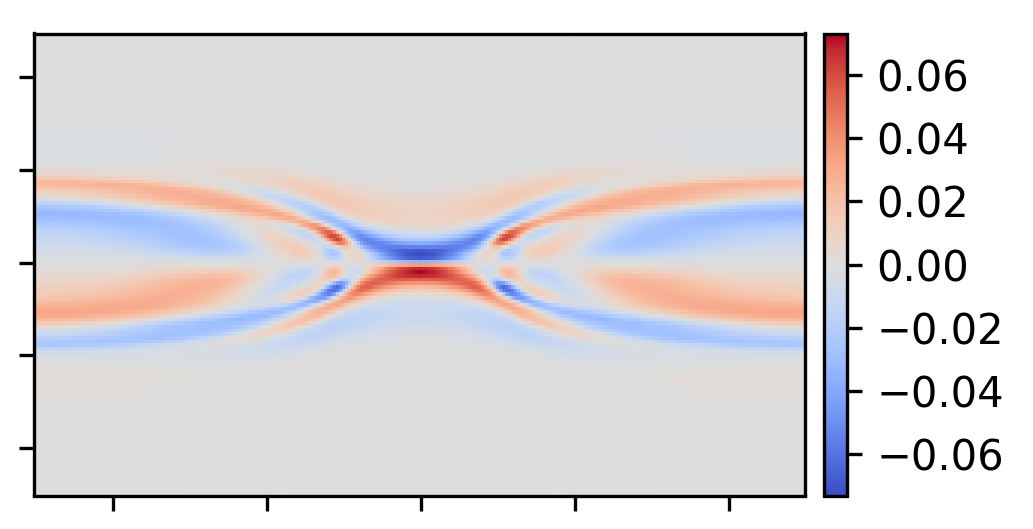} \end{minipage}\\

\begin{minipage}{0.365\textwidth} \includegraphics[height=2.69cm]{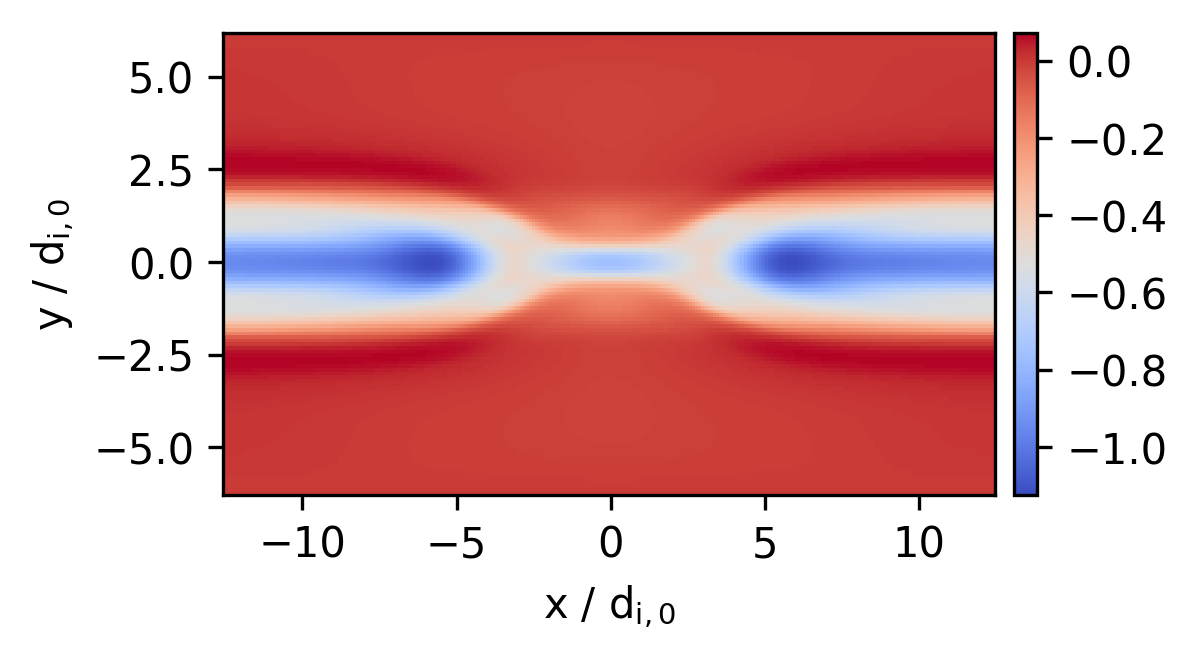} \subcaption{} \end{minipage}
\begin{minipage}{0.305\textwidth} \includegraphics[height=2.69cm]{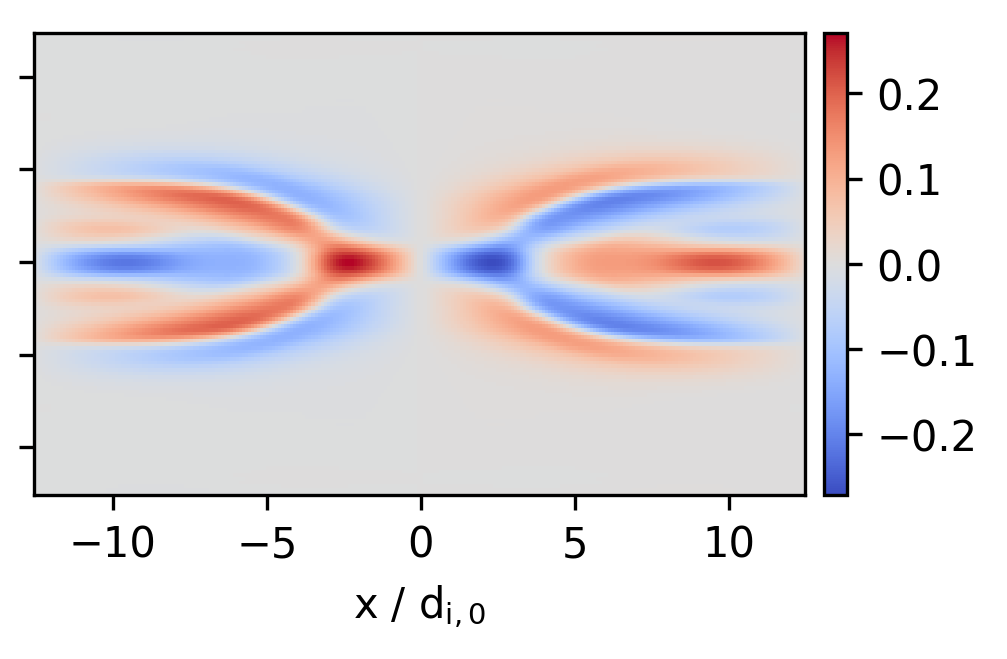} \subcaption{} \end{minipage}
\begin{minipage}{0.305\textwidth} \includegraphics[height=2.69cm]{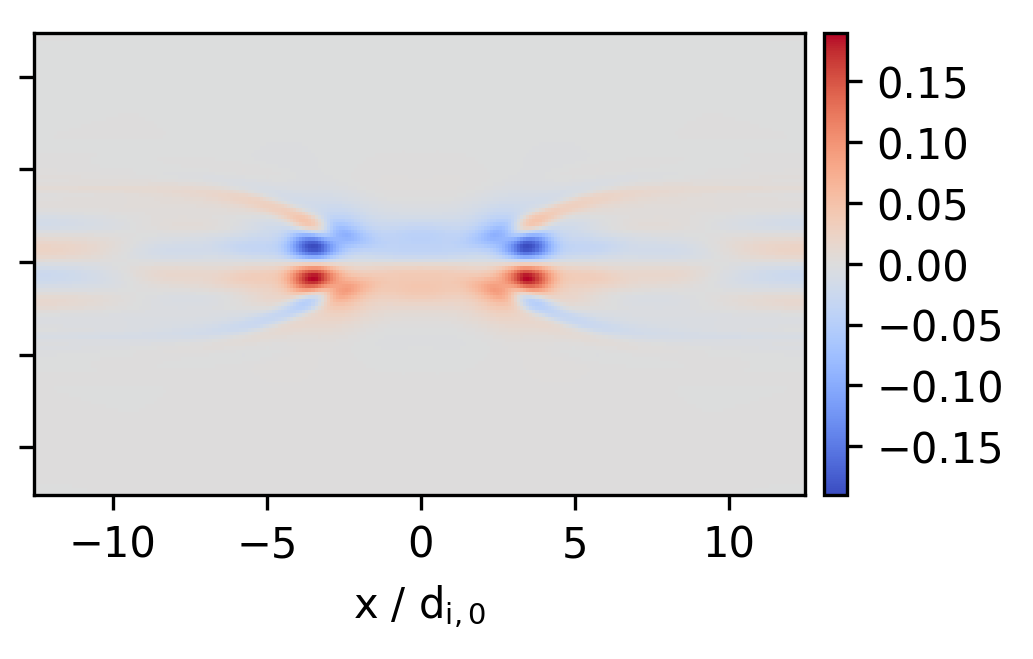} \subcaption{} \end{minipage}

\caption{Vlasov run (first row), scalar-$k$ fluid run (second row) and gradient fluid run (third row).
(a) $j_{z}$ when $\Psi = 1.8$,
(b) $j_{x}$ when $\Psi = 2$,
(c) $P_{xy,e}$ when $\Psi = 2$.\\}
\label{fig:compare_gem}\end{figure}

In the GEM setup the gradient fluid run is compared to both the kinetic Vlasov run and a scalar-$k$ fluid run.
The respective plots can be seen in Fig. \ref{fig:compare_gem}. Snapshots are taken at times where a similar amount of
flux has reconnected since fluid simulations of Harris sheet reconnection usually have a longer
onset than kinetic ones. This is measured by integrating the absolute value of the y-component of the magnetic field, i.e. the reconnected
flux is $\Psi = \int \mathrm{dx}\ |B_{y}|/2$. \\

The scalar-$k$ simulation of the GEM setup that is displayed here was computed with $k_{0,i} = 0.3\ d_{i,0}^{-1}$ and
$k_{0,e} = 5\ d_{i,0}^{-1}$ which gave the best agreement with the Vlasov simulation (cf. Sec. \ref{sec:better_k0}).
Significant improvement can be observed in the run with the gradient closure throughout all parameters.
Also, time development of the gradient run is closer to the Vlasov run.
Fig.\ \ref{fig:compare_gem}a shows that the extremum of the current after reconnection is caught better. Details like
the extrema and the changing sign in Fig.\ \ref{fig:compare_gem}b in the outer areas around the x-axis that the scalar-$k$
run misses are now included. This indicates that the new closure provides a better representation of kinetic effects.
Heat flux change directly influences the pressure tensor, so it is particularly interesting that the agreement with
the kinetic pressure tensor has improved. In Fig.\ \ref{fig:compare_gem}c an example is shown where
the scalar-$k$ closure produces a result significantly different from the Vlasov run while the result from the
gradient closure is very similar.\\

\subsection{WHBG}

\begin{figure}
\begin{minipage}{0.51\textwidth} \includegraphics[height=3.8cm]{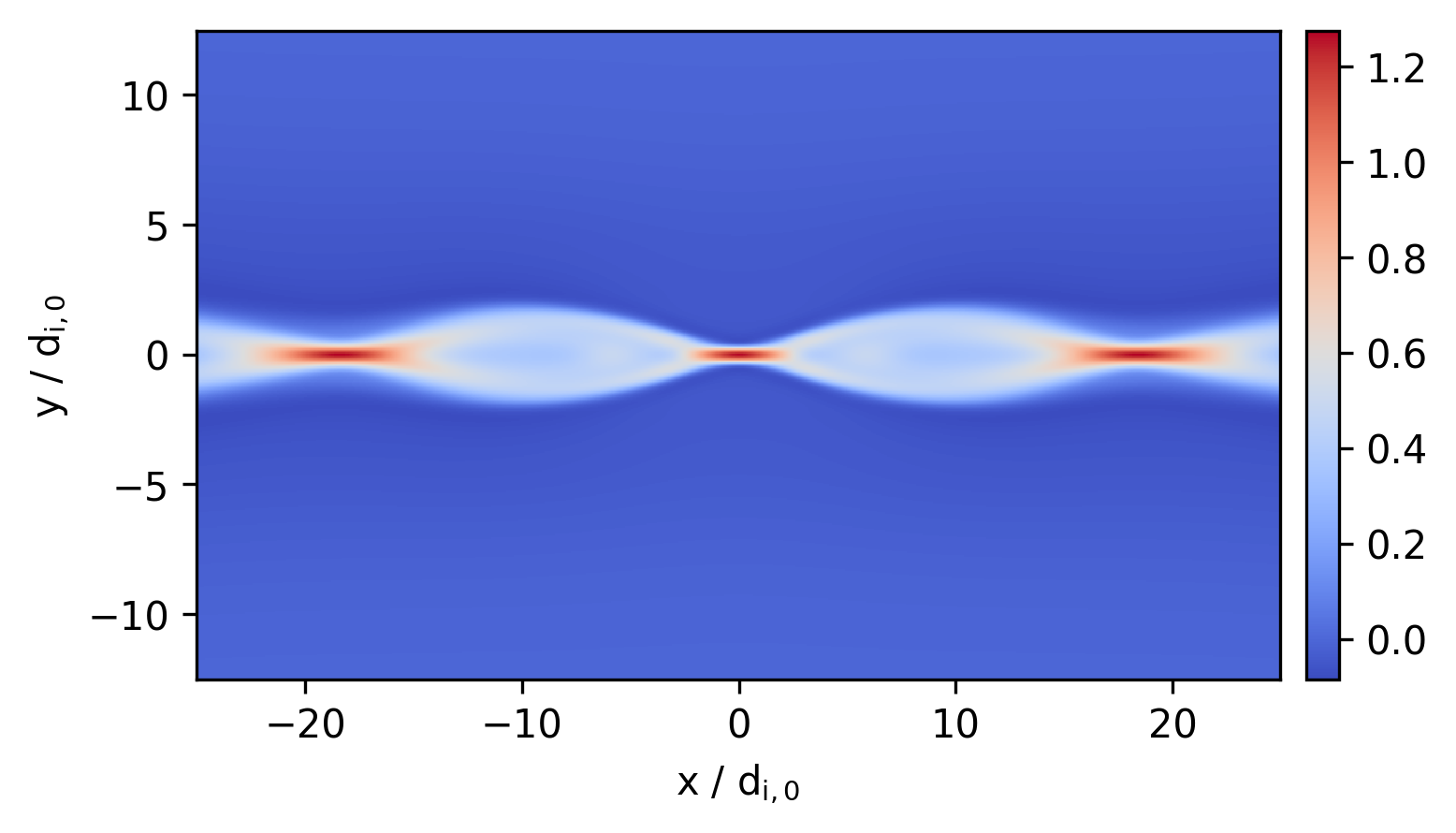} \end{minipage}
\begin{minipage}{0.48\textwidth} \includegraphics[height=3.8cm]{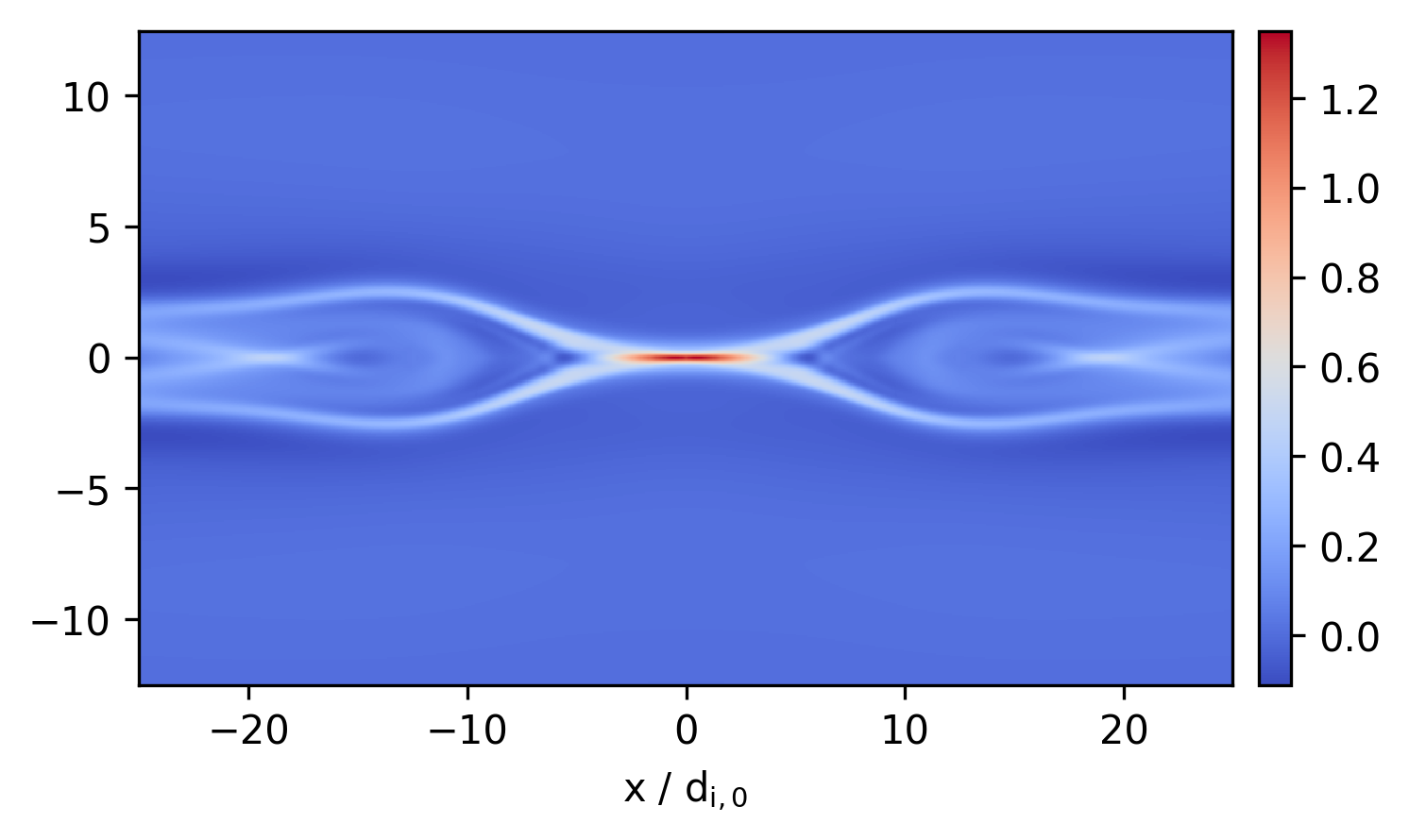} \end{minipage}
\caption{Scalar-$k$ closure (left) and gradient closure (right) in the WHBG setup when $\Psi = 3$.}
\label{fig:compare_whgb}\end{figure}

Originally, the scalar-$k$ closure was tested by \cite{wang-hakim-etal:2015} in the case of reconnection in a larger domain of size
$(100 \times 50)\ d_{i,0}$. They compared the ten moment scalar-$k$ closure to a
kinetic particle in cell (PIC) simulation and found good agreement but some issues as well. Fig.\ \ref{fig:compare_whgb}
shows ten moment runs of the WHBG setup with the scalar-$k$ closure on one hand and the gradient closure
on the other hand with a resolution of $2048\times1024$ cells. There are differences
between our scalar-$k$ run and the plots by Wang et al.\ which might be attributed to the different numerical schemes used,
CWENO here and discontinuous Galerkin in their case \citep{loverich-hakim-shumlak:2011}. Despite the plasmoids
that form in the scalar-$k$ simulation, time development is the same as in Wang et al.'s version (see the next section
for further discussion).
In the gradient run, however, no plasmoids form and the shape is very similar to Wang et al.'s PIC run.
The characteristic wave numbers in the gradient closure were chosen as $k_{0,s} =  \frac{1}{3}\ d_{s,0}^{-1}$.\\

\subsection{Island coalescence}

\begin{figure}
\begin{minipage}{0.53\textwidth} \includegraphics[height=7.5cm]{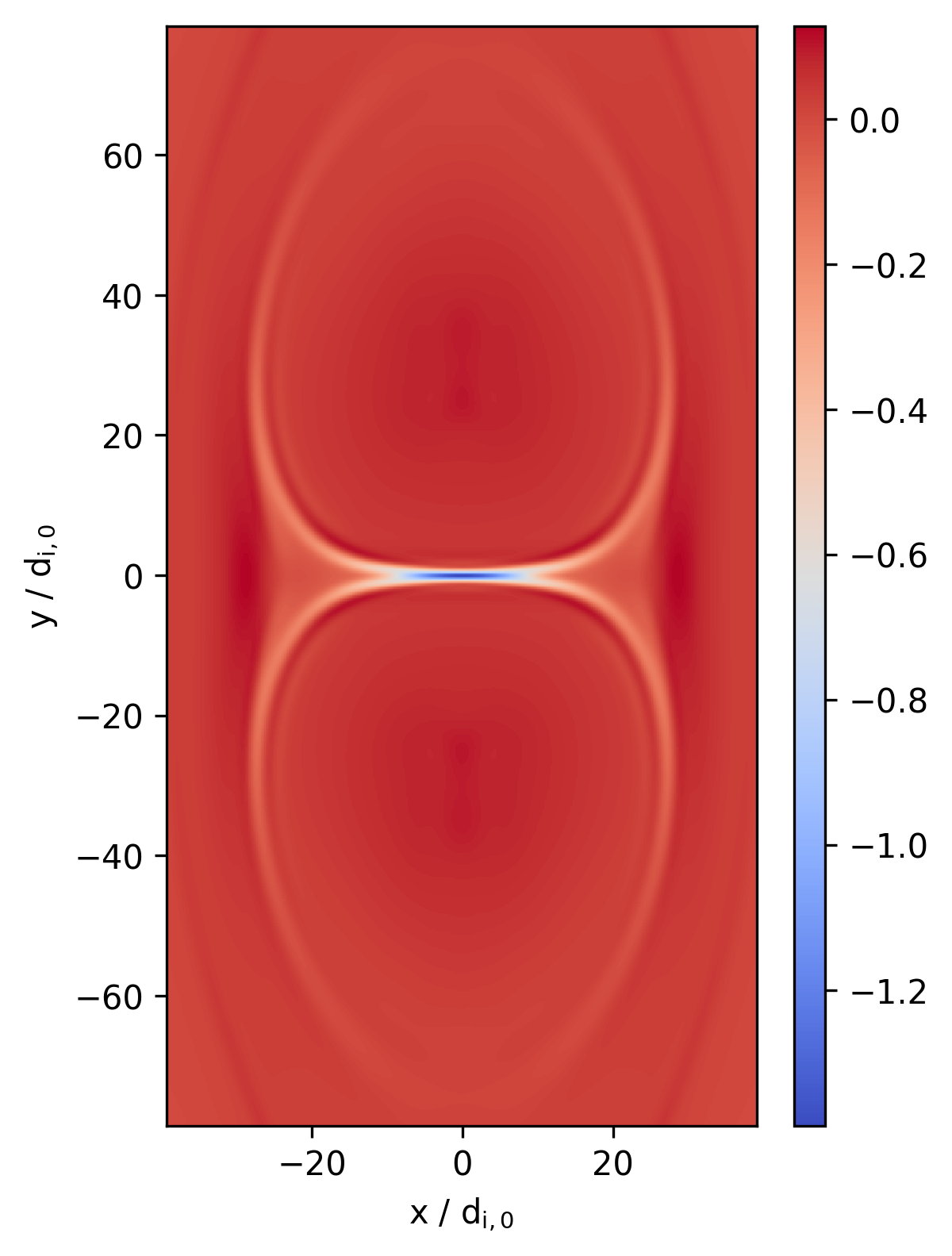} \end{minipage}
\begin{minipage}{0.47\textwidth} \includegraphics[height=7.5cm]{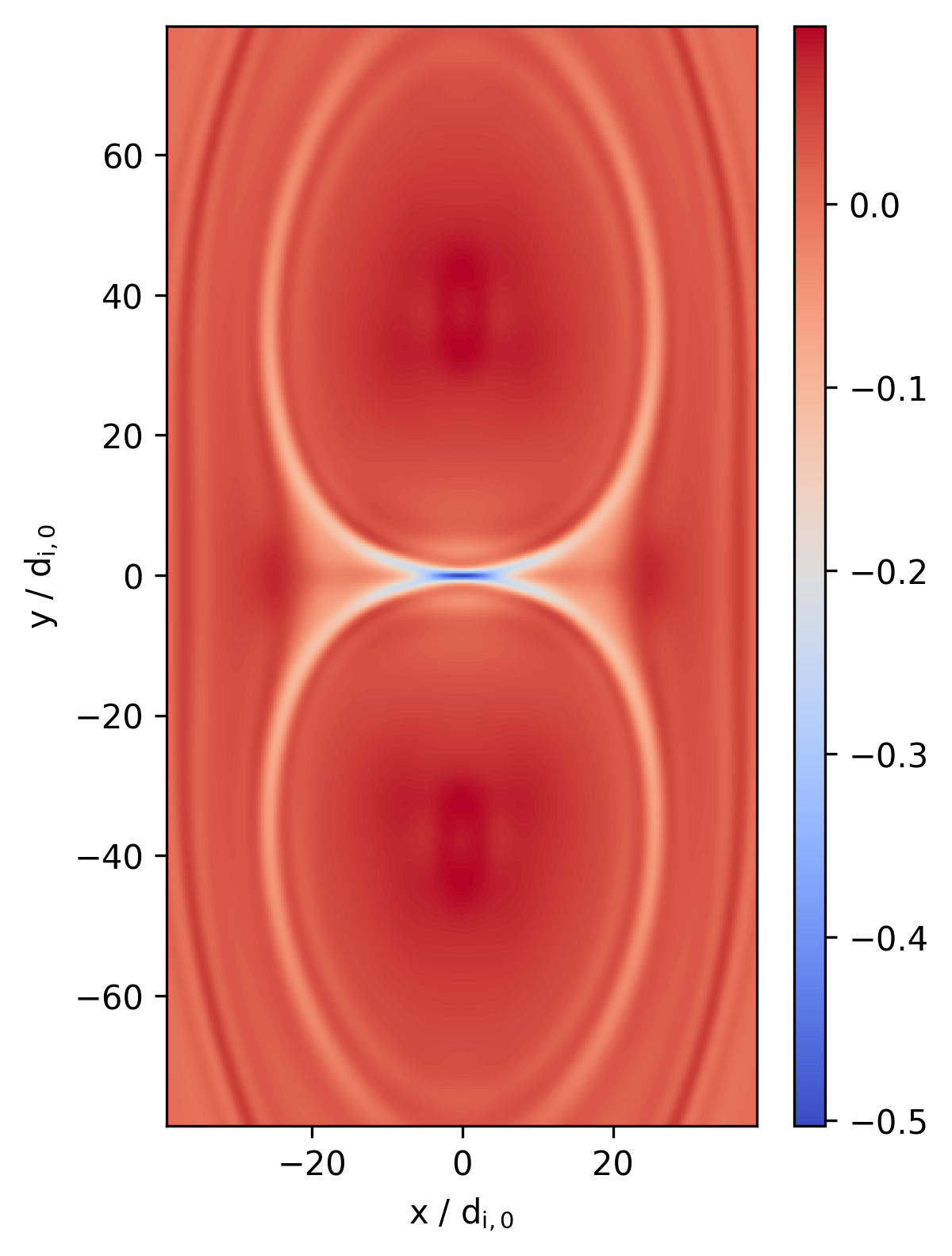} \end{minipage}
\caption{The island coalescence setup at $t = t_{A}$ for $\lambda = 25\ d_{i,0}$. Scalar-$k$ closure (left) and gradient closure (right).}
\label{fig:compare_isl}\end{figure}

\begin{figure}
\begin{minipage}{0.5\textwidth} \includegraphics[height=5cm]{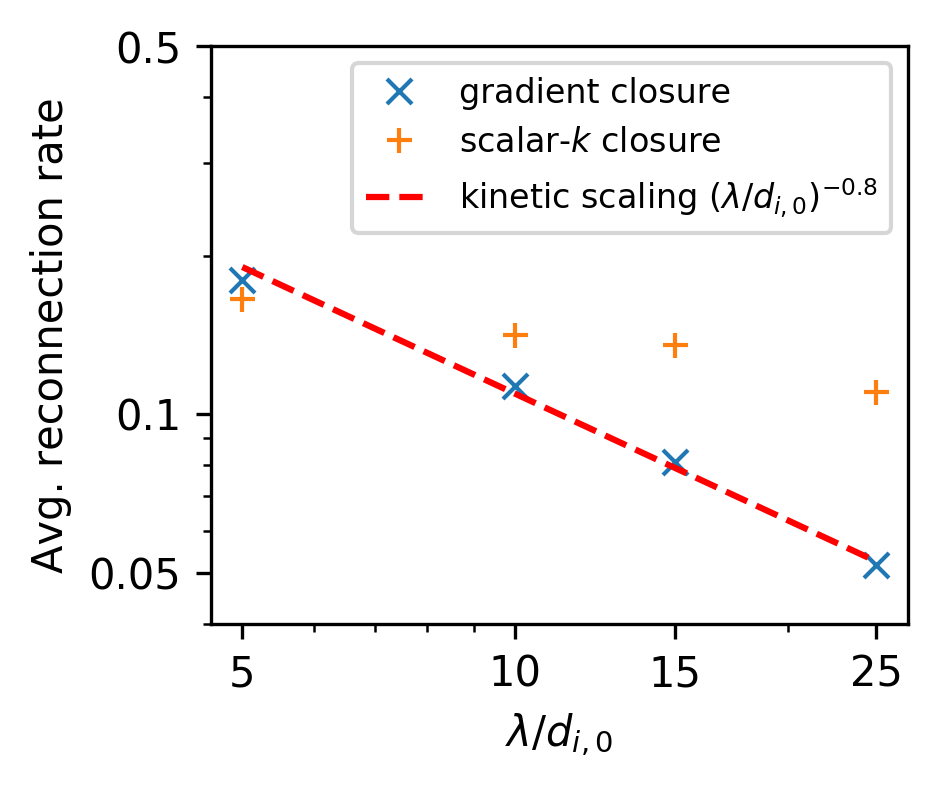} \subcaption{}\end{minipage}
\begin{minipage}{0.5\textwidth} \includegraphics[height=5cm]{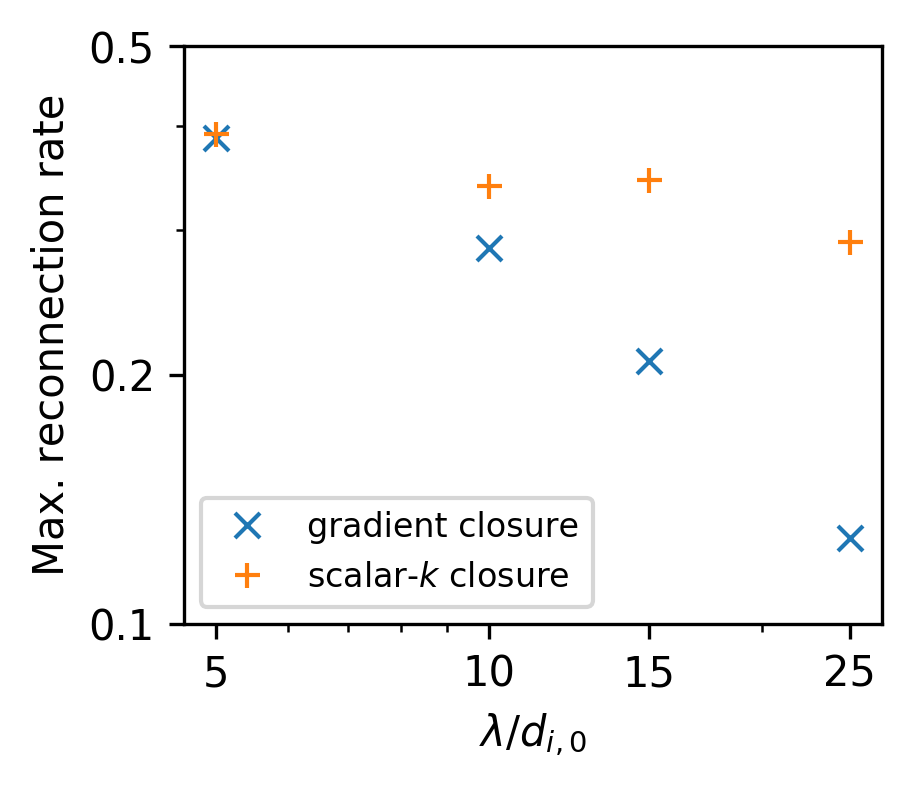} \subcaption{}\end{minipage}\\
\begin{minipage}{0.5\textwidth} \includegraphics[height=5cm]{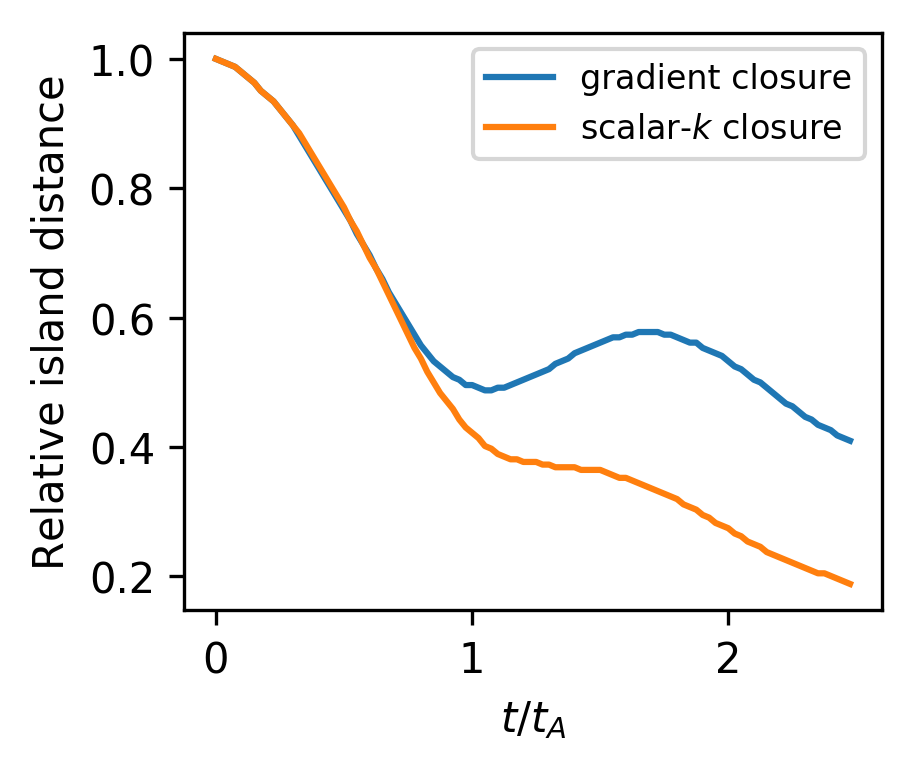} \subcaption{}\end{minipage}
\begin{minipage}{0.5\textwidth} \end{minipage}\\
\caption{Coalescence of magnetic islands. (a) Scaling of the average reconnection rate with the size parameter $\lambda$ for both closures
next to the scaling found by \cite{stanier-daughton-etal:2015} in kinetic PIC simulations which was $\propto (\lambda / d_{i,0})^{-0.8}$. (b) Scaling of the maximum reconnection rate.
(c) Distance of the islands' O-points relative to their initial distance for $\lambda = 15\ d_{i,0}$.}
\label{fig:isl_rec}\end{figure}

The coalescence of islands has been observed in space plasmas and is reported to accelerate electrons to high energies \citep{song-chen-etal:2012}.
Until now no fluid or MHD model was capable of reproducing the kinetic effects in island coalescence well.
\citet{ng-huang-hakim-etal:2015} found good agreement of the scalar-$k$ ten moment model with PIC runs on small spatial scales with
$k_{0,e} = 5\ d_{i,0}^{-1},\ k_{0,i} = 0.3\ d_{i,0}^{-1}$ as the optimal wave number values. Going to larger islands, however,
average reconnection rates decreased according to $(\lambda / d_{i,0})^{-0.2}$ whereas there
was a stronger scaling of $(\lambda / d_{i,0})^{-0.8}$ in kinetic PIC simulations (see also \citet{stanier-daughton-etal:2015}).
There were also further differences from kinetic simulations, e.g.\ islands did not bounce from each other and secondary islands
formed in larger systems.\\

\citet{ng-hakim-etal:2017} proposed a global generalization (see Eq. \ref{eq:ng_global}) of the Hammett-Perkins closure to tensors and tested it in the island coalescence
setup. The generalization is in Fourier space which is computationally expensive but has the advantage that no $k_{0}$ needs to be
chosen. It performed better than the scalar-$k$ closure concerning the average reconnection rates ($\propto (\lambda / d_{i,0})^{-0.45}$)
but did not approach the kinetic scaling. Scaling of maximum reconnection rate did not improve significantly and the other discrepancies
mentioned above remained.\\

We conducted runs of the island coalescence problem with the scalar-k and the gradient closure. Resolutions were chosen so that electron
inertial length $d_{e}$ is resolved. The results of our scalar-k simulations
are very similar to those of \citet{ng-hakim-etal:2017} with a scaling of the average reconnection rate $\propto (\lambda / d_{i,0})^{-0.23}$ and
also matching values for the maximum reconnection rate. In our simulations no secondary islands formed though.
This is particularly interesting because in the WHBG setup Wang et al.\ had no secondary islands and we did (cf.\ the previous section)
while here it is the other way around. Anyway, in both cases the appearance of plasmoids seems to have only minor influence on time development
and reconnection rates.\\

Fluid simulations with the gradient closure and $k_{0,s} = \frac{1}{2} / d_{s,0}$ show the characteristics of kinetic simulations. The average reconnection rate
(average taken from 0 to $1.5\ t_{A}$) is displayed in Fig.\ \ref{fig:isl_rec}a and scales as $(\lambda / d_{i,0})^{-0.73}$ which is almost identical
to the kinetic scaling. Scaling of the maximum reconnection rate is much stronger than with the scalar-$k$ closure as well (Fig.\ \ref{fig:isl_rec}b).
There is no formation of secondary islands. Due to the lower reconnection rates,
islands now bounce as can be seen in Fig.\ \ref{fig:isl_rec}c. The out-of-plane current $j_{z}$ is displayed in Fig.\ \ref{fig:compare_isl} for both
closures. The current sheet and the island's oval form in the gradient simulation are similar to results from kinetic simulations (see the movie in
the supplemental material of \cite{stanier-daughton-etal:2015}).\\

\subsection{Numerics}

The Laplacian in the closure was computed explicitely by means of finite differences. Therefore, instabilities are enhanced and
a smaller time step is needed. Time step restrictions increase with higher resolution. For now, this
has been circumvented by subcycling the computation of the Laplacian. Since the domain has to be split up into blocks
for parallelization, and since boundaries are not exchanged in between the subcycles (for performance reasons), inaccuracies occur
at these borders. Furthermore, the velocities and densities needed to compute pressure from the second moment are not
updated between the subcycles, which has little influence though. A comparison of a subcycled version of the WHBG setup
with one without subcycling shows that globally there is no difference and that the approximation is acceptable when used
thoughtfully. The gradient closure runs displayed in Fig.\ \ref{fig:compare_whgb} and Fig.\ \ref{fig:compare_isl} were computed
with 16 subcycles and a time step identical to the respective scalar-$k$ runs. A more sophisticated solution to the time step
problem is left to future work.\\

%% file: conclusion.tex
\section{Conclusion\label{sec:conclusion}}

Following an analysis of kinetic heat flux data, a closure to the ten moment fluid equations
is presented which approximates the heat flux tensor as

\begin{equation}
\partial_{m} \mathrm{Q}_{ijm} = -\frac{v_{t}}{|k_{0,s}|}\ \nabla^{2}\ (\mathrm{P}_{ij} - p \delta_{ij})
\label{eq:d2p_final}\end{equation}

with the free parameter $k_{0,s}$ (a typical wave number) and $p = (\mathrm{P}_{xx} + \mathrm{P}_{yy} + \mathrm{P}_{zz})/3$.
Suitable values for $k_{0,s}$ in magnetic reconnection are $3 / d_{s,0}$ in the GEM setup,
$\frac{1}{3} / d_{s,0}$ in the WHBG setup and $\frac{1}{2} / d_{s,0}$ in island coalescence.\\

The derivation of Eq.\ \ref{eq:d2p_final} used findings of \cite{hammett-perkins:1990} and \cite{wang-hakim-etal:2015}.
The approximations made were motivated by a test of the original one-dimensional Hammett-Perkins
approach along magnetic field lines.
The new closure was tested in three different reconnection setups and the results agreed well with kinetic Vlasov and PIC
simulations in all cases. Good results were achieved in the coalescence of magnetic islands where fluid models were unsuccessful before.
Including the pressure gradient is supposed to improve the modeling of kinetic effects like Landau damping so that the fluid equations
can replace expensive kinetic computations. That way simulations of large spatial scales like Earth's magnetotail become within reach.\\

Future work includes further investigation of the free parameter because currently it has to be determined from experiments
and the comparison with kinetic simulations. The focus should be on the effect of different setups since the free parameter appears
to be specific to the respective problem. Another approach would be to couple the fluid code to Vlasov computations in order to
adaptively adjust the free parameter. From a technical point of view, elaborate solutions to the time step restrictions caused
by the Laplacian would be desirable.\\

%% file: acknowledgments.tex
FAR appreciated the helpful discussions with Simon Lautenbach.
Computations were conducted on the Davinci cluster at TP1 Plasma Research Department and on the JURECA cluster at
J\"ulich Supercomputing Center under the project number HBO43.\\